# The multifunctionality of lanthanum-strontium cobaltite nanopowder: high-pressure magnetic and excellent electrocatalytic properties for OER

Hanlin Yu[a], N.A. Liedienov[a,b,*], I.V. Zatovsky[c], D.S. Butenko[d,e,*], I.V. Fesych[f], Wei Xu[g], Songchun Rui[h], Quanjun Li[a], Bingbing Liu[a], A.V. Pashchenko[a,c,i], G.G. Levchenko[a,c,*]

[a]*State Key Laboratory of Superhard Materials, International Center of Future Science, Jilin University, 130012 Changchun, P.R. China*
[b]*Donetsk Institute for Physics and Engineering named after O.O. Galkin, NASU, 03028 Kyiv, Ukraine*
[c]*F.D. Ovcharenko Institute of Biocolloidal Chemistry, NASU, 03142 Kyiv, Ukraine*
[d]*Shenzhen Key Laboratory of Solid State Batteries, Southern University of Science and Technology, Shenzhen 518055, P.R. China*
[e]*Academy for Advanced Interdisciplinary Studies, Southern University of Science and Technology, Shenzhen 518055, P.R. China*
[f]*Taras Shevchenko National University of Kyiv, 01030 Kyiv, Ukraine*
[g]*State Key Laboratory of Inorganic Synthesis and Preparative Chemistry, College of Chemistry, Jilin University, Changchun, 130012, P.R. China*
[h]*Baicheng Normal University, 137099 Baicheng, China*
[i]*Institute of Magnetism NASU and MESU, 03142 Kyiv, Ukraine*

*Corresponding author
E-mail address:     nikita.ledenev.ssp@gmail.com (N.A. Liedienov)
                    debut98@ukr.net (D.S. Butenko)
                    g-levch@ukr.net (G.G. Levchenko)

**Abstract**

Simultaneous study of magnetic and electrocatalytic properties of cobaltites under extreme conditions expands understanding of physical and chemical processes proceeding in them with the possibility of their further practical application. Therefore, $La_{0.6}Sr_{0.4}CoO_3$ (LSCO) nanopowders have been synthesized at different annealing temperatures $t_{ann}$ = 850, 875, 900 °C, and their multifunctional properties have been studied comprehensively. As $t_{ann}$ increases, the rhombohedral (sp. gr. $R\bar{3}c$) perovskite structure of the LSCO becomes more single-phase, whereas average particle size and dispersion grow. Co ions are in mixed valence states, including major $Co^{3+}$ and $Co^{4+}$ components. It has been found that the LSCO-900 shows two main Curie temperatures, $T_{C1}$ and $T_{C2}$, associated with a particle size distribution. As an external hydrostatic pressure $P$ increases, average $<T_{C1}>$ and $<T_{C2}>$ increase from 253 and 175 K under ambient pressure to 268 and 180 K under $P$ = 0.8 GPa, respectively. At the same time, the antiferromagnetic temperature $T_{AFM}$ and blocking temperature $T_B$ also increase from 145 and 169 K to 158 and 170 K, respectively. The increment of $<dT_C/dP>$ for the smaller and bigger particles is sufficiently high and equals 10 and 13 K/GPa, respectively. The magnetocaloric effect in the LSCO-900 nanopowder is relatively weak but with an extremely wide peak $\delta T_{FWHM}$ > 50 K that makes this composition interesting to be used as one of the components of the composite expanding its working temperature window. Moreover, all LSCO samples showed excellent electrocatalytic performance for the overall water splitting (OER) process (overpotentials only 265–285 mV at a current density of 10 mA cm$^{-2}$) with minimal $\eta_{10}$ for LSCO-900. Based on the XPS data, it was found that the formation of a dense amorphous layer on the surface of the particles ensures high stability as a catalyst (at least 24 h) during electrolysis in 1 M KOH electrolyte.

# 1. Introduction

Interest in lanthanum-doped strontium $La_{1-x}Sr_xCoO_3$ cobaltites is constantly increasing due to their physical and chemical multifunctionality [1-5]. They demonstrate unique magnetic and transport properties [6, 7], as well as charge, spin, and orbital ordering [8], spin-state transitions [9], electromagnetic phase separation [10], and metal-insulator transitions [11] with strong mutual spin, orbital, charge, and lattice coupling [12]. It makes them ferromagnetic (FM), ferroelectric, and electrocatalytically active with high electrode performance and giant magnetoresistance effect [13, 14]. Moreover, they show high ionic and electronic conductivity, good catalytic properties, and high stability [15], especially because of the low mobility of Sr on the crystal surface and the formation of $Co^{4+}$ ions and oxygen vacancies [16]. All these attract great scientific interest to cobaltites and make them promising candidates for their practical application as electrode materials for high-temperature solid oxide fuel cells, catalysts, current and magnetic field sensors, gas sensors, etc. [17-23]. However, there are still unclear and unstudied physical and chemical processes as the surface-dependent reasons for obtaining high electrode performance and the influence of high external hydrostatic pressure on the magnetic phase transitions and magnetocaloric effect (MCE).

Original $LaCoO_3$ has an antiferromagnetic (AFM) order, and dilution with divalent $Sr^{2+}$ up to $x = 0.4$ enhance its FM order making this $La_{0.6}Sr_{0.4}CoO_3$ compound fully FM with the highest Curie temperature $T_C \approx 230$ K and the closest to the room temperature among other $La_{1-x}Sr_xCoO_3$ compositions[11]. However, the $T_C$ can significantly differ and be 246 K [7, 24] or even 321 K [25] for the same $x = 0.4$. Moreover, besides classical FM double-exchange (DE) between $Co^{3+}$ and $Co^{4+}$ ions, the AFM superexchange interactions $Co^{3+}$–$Co^{3+}$ and $Co^{4+}$–$Co^{4+}$, and phase separation can coexist [26] that makes this $La_{0.6}Sr_{0.4}CoO_3$ system more complicated along with the sophisticated spin states distribution. Additionally, there is no data about particle size distribution's influence on the magnetic properties, Curie temperature, and MCE of the $La_{0.6}Sr_{0.4}CoO_3$ nanopowders. The effect of high hydrostatic pressure on their functional properties is of particular interest. External pressure should modify Co–O distance and Co–O–Co angle and, as a result, change exchange interactions, phase transition temperatures, degree of magnetic uniformity, and MCE parameters. All these issues and unspecified moments need to be clarified.

Among the various functional properties of materials-based perovskite oxides, including 3$d$-metals (Mn, Fe, Co, Ni), a promising electrocatalytic activity toward oxygen evolution should also be noted [27]. It fully applies to Ln- and Co-containing complex oxides of the $ABO_3$-type, where for perovskites, $A$-positions usually correspond to a rare-earth or alkaline earth element, and $B$ is commonly a transition metal or a combination thereof [28-37]. In general, it is considered that the oxygen-catalytic activity of perovskites is influenced by both types of metal in the $A$- and $B$-sites [38]. However, the high electrocatalytic activity of perovskites for overall water splitting (OER) in an alkaline medium is primarily the result of the redox behavior of the hybridization of transition metal 3$d$ and oxygen 2$p$ orbits [39]. Numerical studies of perovskites as electrocatalysts revealed that, in addition to the chemical composition, activity in oxygen electrocatalysis reactions is also affected by a significant number of factors, namely: method and conditions of preparation of the material (annealing at different temperatures and in a controlled atmosphere can cause different types of defects on the surface) [31, 40]; particle size, porosity, and shape (affects active surface area) [29-31, 41]; doping [28, 32, 34, 42, 43]; amorphization of the surface [44]. Moreover, due to electrochemical processes or an aggressive alkaline electrolyte, complex oxides, including perovskites, can be transformed *in situ* to other compounds [45-47], directly affecting the electrocatalytic activity during long-term electrolysis. Accordingly, when studying perovskites as electrocatalysts for OER, evaluating their primary activity and conducting research after electrochemical testing is crucial.

Additionally, studying electrochemical activity and physical properties directly on the same samples is very important since the obtained data supplements each other and allows us to make more correct conclusions. Moreover, knowledge of the behavior of other physical properties, such as temperature changes of magnetic phase transitions and magnetoactive phenomena (MCE or magnetoresistance effect) under changing internal and external factors, is essential as well when it is necessary to change the conditions of electrocatalytic applications.

Thus, in this paper, the cobaltite $La_{0.6}Sr_{0.4}CoO_3$ nanopowders have been obtained under different annealing temperatures $t_{ann}$ = 850, 875, and 900 °C and their structure, morphology, particle size distribution, as well as magnetic, magnetocaloric, and electrocatalytic properties have been studied comprehensively. The most single-phase $La_{0.6}Sr_{0.4}CoO_3$ nanopowder with $t_{ann}$ = 900 °C has been selected to study the phase transitions and magnetic properties under high hydrostatic pressure up to $P \approx 0.8$ GPa. The same sample showed excellent electrocatalytic activity and high stability for OER in alkaline electrolytes.

## 2. Experimental section

### 2.1. Sample preparation

Cobaltite $La_{0.6}Sr_{0.4}CoO_3$ (LSCO) nanopowders were obtained using the sol-gel method from the initial $La(NO_3)_3 \cdot 6H_2O$ (99.9% trace metals basis, Sigma–Aldrich), $Sr(NO_3)_2$ (ACS reagent, ≥ 99.0%, Sigma–Aldrich) and $Co(NO_3)_2 \cdot 6H_2O$ (reagent grade, 98%, Sigma-Aldrich) reagents. The calculated amount of nitrates and citric acid (the ratio of the molar sum of metals and $H_3Cit \cdot H_2O$ was 1:1) were dissolved in deionized water. The solution was evaporated to form a homogeneous gel, which was dehydrated and heated from 200 to 500 °C (the heating rate equals 50 °C·h$^{-1}$). The obtained powder was ground in an agate mortar and annealed at temperatures $t_{ann}$ = 850, 875, and 900 °C for 10 h at each temperature. Finally, three samples obtained at different $t_{ann}$ were named LSCO-850, LSCO-875, and LSCO-900 for further investigation.

### 2.2. Characterization

The phase composition and size of coherent scattering regions of the samples were examined using the X-ray diffraction (XRD) method on a Shimadzu LabX XRD-6000 diffractometer in Cu$K_{\alpha 1}$-radiation ($\lambda$ = 0.15406 nm) before electrocatalysis and X-ray diffractometer MicroMax-007 HF (Rigaku, Japan) in Mo-$K_{\alpha}$ radiation ($\lambda$ = 0.071146 nm) after electrocatalysis, both at room temperature. The database JCPDS PDF-2 was used for defining the phase composition. The refinement of crystal structures was carried out with Rietveld analysis [48] using FullProf software [49].

The morphology and size of particles were determined using the transmitting electron microscope (TEM) method on a JEM-2200FS Transmission Electron Microscope and scanning electron microscope (SEM) method on a FEI MAGELLAN 400 and Bruker XFlash 6|60 in the Hitachi Regulus 8100. The distribution function of nanoparticle size was obtained by analyzing SEM and TEM images using Nano Measure 1.2.5 software [45]. Additionally, the chemical composition of the samples was clarified using energy-dispersive X-ray spectroscopy (EDS) mode.

X-ray photoelectron spectroscopy (XPS) measurements were performed on an ESCALAB 250 X-ray photoelectron spectrometer. The XPS spectra were fitted and analyzed using OriginPro 2018 software. The background of the X-ray photoelectronic lines was cut using the Shirley method. The spectra were excited using monochromatized Al-Kα radiation. The state of the surfaces was monitored through the C1s line, allowing calibration of the energy scales for all spectra. The C1s line binding energy was approximately 285 eV.

Magnetic measurements were carried out using a Quantum Design SQUID MPMS 3 in a temperature range from 2 to 300 K and in a magnetic field up to 30 kOe. The temperature dependences of magnetization $M(T)$ under $H$ = 50 Oe were carried out in two regimes: zero-field cooling (ZFC) and field cooling (FC). The heating and cooling of samples were carried out with a constant rate of 1 K·min$^{-1}$. The MCE was determined using magnetization isotherms $M(H)$ near Curie temperature $T_C$ with a $\Delta T$ = 2 K step. Each isotherm $M(H)$ was measured with an increase in the magnetic field $H$ from 0 to 10 kOe with a step of $\Delta H$ = 100 Oe. Before each measurement of isotherms, magnetic nanopowder was demagnetized. Additionally, magnetic measurements under high pressure up to $P \approx$ 0.8 GPa were performed using a piston-type pressure cell made of Ni–Cr–Al alloy [50, 51]. Silicon oil of low viscosity was used as a pressure-transmitting medium. The pressure inside was measured using the pressure dependence of the superconducting transition temperature of high-purity lead.

**2.3. Electrodes fabrication and electrochemical measurements**

To fabricate the working electrodes, mixtures of LSCO samples, Super P (conductive carbon black), and PVDF (polyvinylidene fluoride) were prepared in a weight ratio of 85:10:5, respectively. Next, NMP (N-methylpyrrolidone) was added to the mixture. After that, the liquid suspension was applied onto a carbon fiber (HCP331N) with a working area of 1 cm$^2$ and dried at 60 °C for 24 h in a vacuum oven. Each electrode contained approximately about 30 mg of an electroactive material. The benchmark RuO$_2$/C electrode for investigating OER was made using a similar preparation procedure.

The electrochemical tests were conducted using the electrochemical workstation (CHI 760E, Shanghai, China). A typical three-electrode setup included an LSCO working electrode, a commercial Hg/HgO reference electrode, and a platinum foil counter electrode in 1M KOH (pH = 14). The linear sweep voltammetry (LSV) curves were collected at a scan rate of 1 mV·s$^{-1}$ with iR-compensation. Electrochemical impedance spectroscopy (EIS) was obtained in a frequency range from 0.1 to 100 kHz with an amplitude of 5 mV at the open circuit potential. Chronopotentiometry (CP) was carried out under a constant current density of 10 mA·cm$^{-2}$ for 24 h. All the potentials versus Hg/HgO were converted to versus reversible hydrogen electrode (RHE): ($E_{RHE} = E_{Hg/HgO} + 0.098 + 0.059 \times pH$).

**3. Results and discussion**

**3.1. Structural properties**

XRD patterns of the LSCO nanopowders are presented in Fig. 1. The main crystalline phase is a perovskite crystal structure with a rhombohedral $R\bar{3}c$ space group. Some of the second-phase was also detected as SrCo$_{0.78}$O$_{2.48}$, which decreases with increasing $t_{ann}$ from 17.4% for LSCO-850 to 14.8% for LSCO-875 and 8% for LSCO-900. It should be noted that this impurity with increased content of cationic "Co$_{0.78}$" and anionic "O$_{2.48}$" vacancies does not influence the magnetic properties of the main LSCO phase near its high Curie temperature $T_C \approx 250$ K (see the following Section 3.2.) because the closest SrCoO$_{2.5}$ composition to our impurity is AFM with high Néel temperature $T_N = 570$ K [52]. Moreover, the less content of oxygen and cobalt, the much lower $T_C$ compared with our $T_C \approx 250$ K and much weaker FM interactions [8, 53-56]. The main lattice parameters are listed in Table 1 and are consistent with other literature data [24, 57]. With increasing $t_{ann}$, a unit cell volume $V$ is increased non-monotonically, indicating the successive replacement of smaller $La^{3+}$ cation ($r_{La^{3+}}$ = 0.136 nm) by bigger $Sr^{2+}$ cation ($r_{Sr^{2+}}$ = 0.144 nm) in 12-fold oxygen coordination [58]. At the same time, the average size of coherent scattering regions $D_{XRD}$, determined via the Scherrer equation [59], increases with increasing $t_{ann}$ from 48 nm for the LSCO-850 to 54 nm for the LSCO-875 and 60 nm for the LSCO-900 (Table 2 and ESI 1).

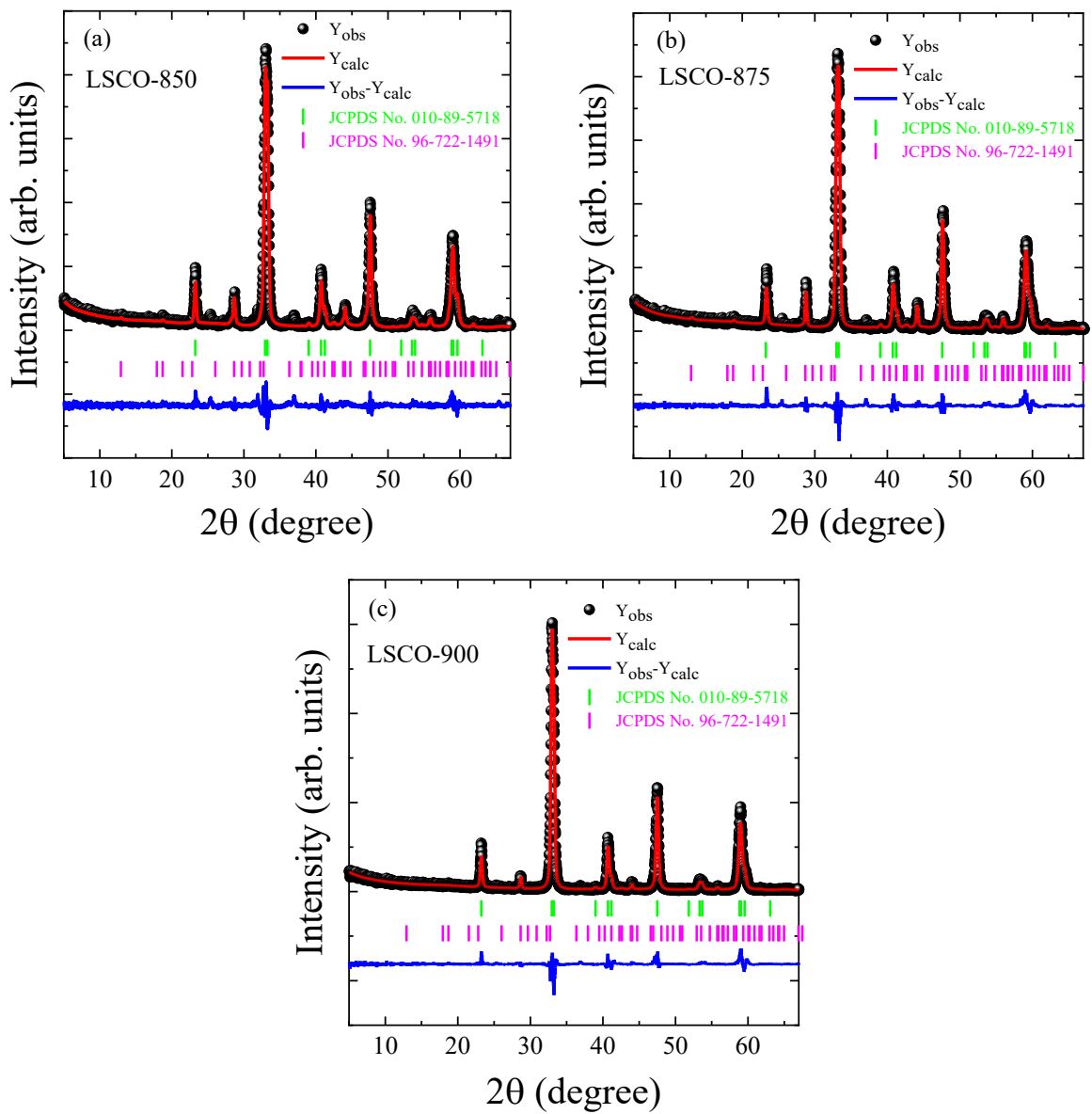

**Fig. 1.** XRD patterns of the LSCO measured at room temperature and fitted by the Rietveld method. The experimental (black cycles) and calculated (red up curve) values and a difference curve (blue bottom line) normalized to a statistical error are presented. Vertical bars are the calculated positions of diffraction peaks corresponding to the main rhombohedral $R\bar{3}c$ perovskite structure and rhombohedral $R32$ impurity $SrCo_{0.78}O_{2.48}$ in the LSCO-850, LSCO-875, and LSCO-900.

**Table 1**
**Rietveld refinement crystallographic parameters of the LSCO nanopowders at different $t_{ann}$**

| $t_{ann}$ (°C) | 850 | | 875 | | 900 | |
|---|---|---|---|---|---|---|
| Phase composition | $La_{0.6}Sr_{0.4}CoO_3$ 82.6% | $SrCo_{0.78}O_{2.48}$ 17.4% | $La_{0.6}Sr_{0.4}CoO_3$ 85.2% | $SrCo_{0.78}O_{2.48}$ 14.8% | $La_{0.6}Sr_{0.4}CoO_3$ 92.0% | $SrCo_{0.78}O_{2.48}$ 8.0% |
| Space group | $R\bar{3}c$ (No. 167) | $R32$ (No. 155) | $R\bar{3}c$ (No. 167) | $R32$ (No. 155) | $R\bar{3}c$ (No. 167) | $R32$ (No. 155) |
| $a$ (Å) | 5.43171(24) | 9.46930(176) | 5.42651(22) | 9.46644(123) | 5.43219(17) | 9.47809(280) |
| $b$ (Å) | 5.43171(24) | 9.46930(176) | 5.42651(22) | 9.46644(123) | 5.43219(17) | 9.47809(280) |
| $c$ (Å) | 13.12327(97) | 12.39900(378) | 13.12711(86) | 12.36905(277) | 13.14732(83) | 12.37605(411) |
| $V$ (Å$^3$) | 335.310(32) | 962.837(387) | 334.766(29) | 959.931(278) | 335.984(26) | 962.840(514) |
| $Z$ | 6 | 18 | 6 | 18 | 6 | 18 |
| $\rho$ (g/cm$^3$) | 6.695 | 5.379 | 6.708 | 5.395 | 6.676 | 5.379 |
| $R_p$ (%) | 10.9 | | 10.4 | | 10.3 | |
| $R_{wp}$ (%) | 15.2 | | 14.1 | | 13.6 | |
| $R_{exp}$ (%) | 9.76 | | 10.05 | | 9.94 | |
| $\chi^2$ (%) | 2.43 | | 1.98 | | 1.87 | |

The morphology and particle size distribution of the LSCO were studied using SEM and TEM data (see Figs. 2 and 3). All LSCO nanopowders demonstrate well-developed structural nature with a spherical-like shape of nanoparticles. An average particle size $D$ and their distribution (see Table 2) were defined using a detailed approach described in our previous work [45]. With increasing $t_{ann}$, the particle size is increased from $D_{SEM} = 74$ nm and $D_{TEM} = 39$ nm for the LSCO-850 to 75 and 44 nm for the LSCO-875 and 79 and 47 nm for the LSCO-900, which correlates with the XRD data. The differences between SEM and TEM data may be associated with insufficient datasets by size. It should be noted that a dispersion $\sigma$ is relatively high, indicating the existence of particles with a big difference in their sizes. Additionally, the chemical composition of all LSCO is approximately confirmed as 0.6:0.4:1 according to the EDS data (see Fig. 2 (b, d, f)).

**Table 2**
**Particle size distribution based on the XRD, SEM, and TEM data for LSCO at different $t_{ann}$**

| $t_{ann}$ (°C) | XRD | SEM | | | | TEM | | | |
|---|---|---|---|---|---|---|---|---|---|
| | $D_{XRD}$ (nm) | $D_{SEM}$ (nm) | Function | $\sigma$ | $R^2$ | $D_{TEM}$ (nm) | Function | $\sigma$ | $R^2$ |
| 850 | 48±1 | 74±3 | Lorentz | 28±9 | 0.62297 | 39±2 | LogNormal | 32±6 | 0.87573 |
| 875 | 54±1 | 75±2 | Gaussian | 54±5 | 0.96482 | 44±1 | LogNormal | 28±2 | 0.98392 |
| 900 | 60±1 | 79±1 | Gaussian | 63±3 | 0.98155 | 47±1 | LogNormal | 13±2 | 0.97304 |

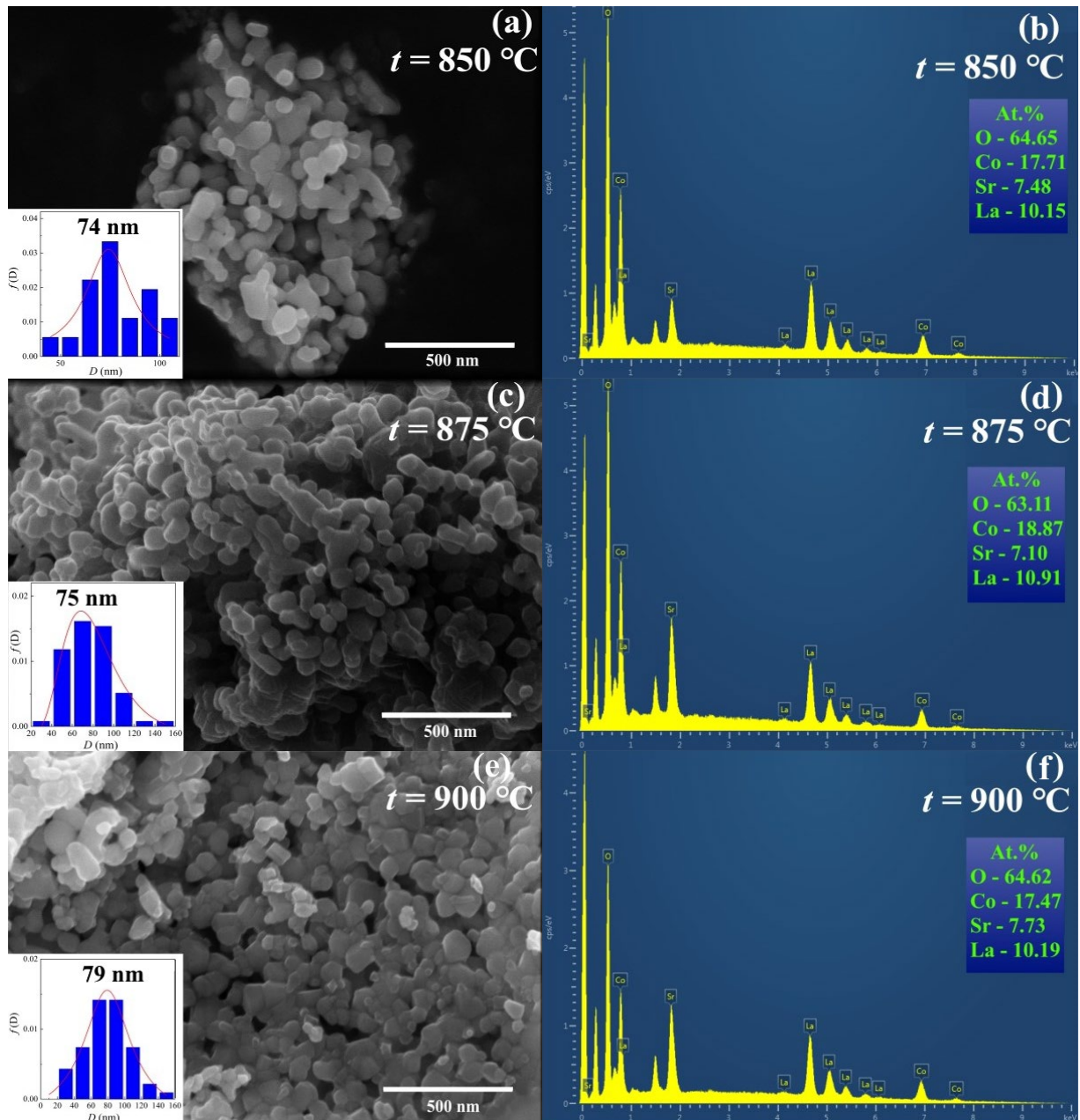

**Fig. 2.** SEM and EDS images of the LSCO compositions obtained at $t_{ann}$ = 850 (a, b), 875 (c, d), and 900 °C (e, f). Insets show particles' distribution with an average particle size $D_{SEM}$.

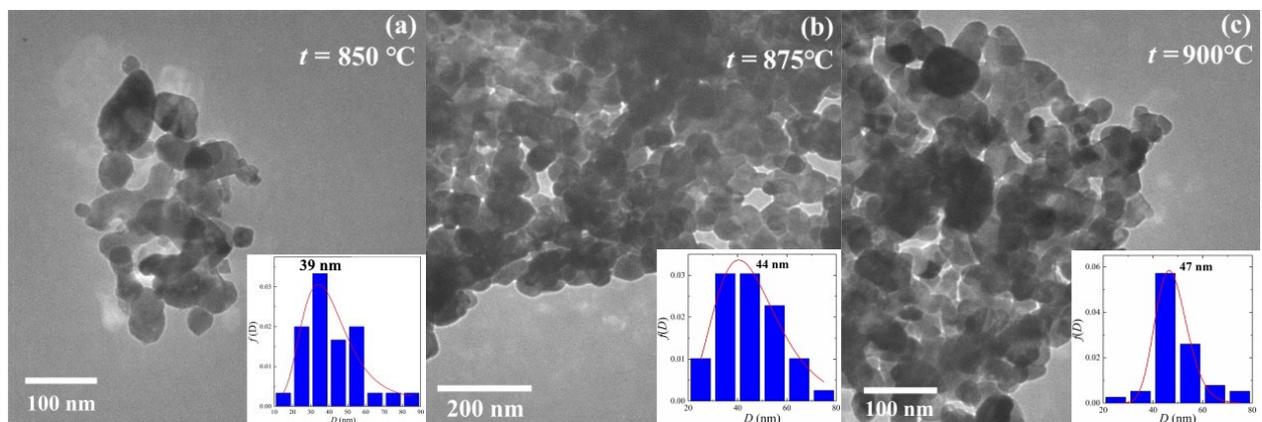

**Fig. 3.** TEM images of the LSCO nanopowders obtained at $t_{ann}$ = 850 (a), 875 (b), and 900 °C (c). Insets show particles' distribution with an average particle size $D_{TEM}$.

## 3.2. Magnetic properties

The field dependences of magnetization $M(H)$ for the LSCO-900 at 2, 77, and 300 K are presented in Fig. 4. The LSCO-900 is in FM at 2 and 77 K and PM at 300 K states. At 2 K, the spontaneous magnetization $M_S$ achieves 10.5 emu/g, and there is no saturation even under magnetic field 3 T that indicates competing between FM DE $Co^{3+}$–$Co^{4+}$ and AFM superexchange $Co^{3+}$–$Co^{3+}$ and $Co^{4+}$–$Co^{4+}$ interactions [60]. At the same time, Co ions can be in high-spin (HS), low-spin (LS), and/or even intermediate-spin (IS) states [12]. Based on the experimentally defined FM moment $\mu_{FM}^{exp}$ = 0.42$\mu_B$ and other data [6, 7, 11, 53], it can be concluded that the most probable spin state for $Co^{3+}$ and $Co^{4+}$ ions are HS ($t_{2g}^4 e_g^2$) with $S = 4/2$ and LS ($t_{2g}^5 e_g^0$) with $S = 1/2$ states, respectively. Theoretical calculation gives $\mu_{FM}^{theor}$ = 2.8$\mu_B$. The big difference between theoretical and experimental data may be associated with competing for different magnetic interactions and the presence of oxygen vacancies [7, 50, 61, 62]. The main magnetic parameters are listed in Table 3. It should be noted that the LSCO-900 is a ferromagnet with a strong coercivity $H_C$ = 4.3 kOe and residual magnetization $M_r$ = 7.2 emu/g at 2 K (see the inset of Fig. 4).

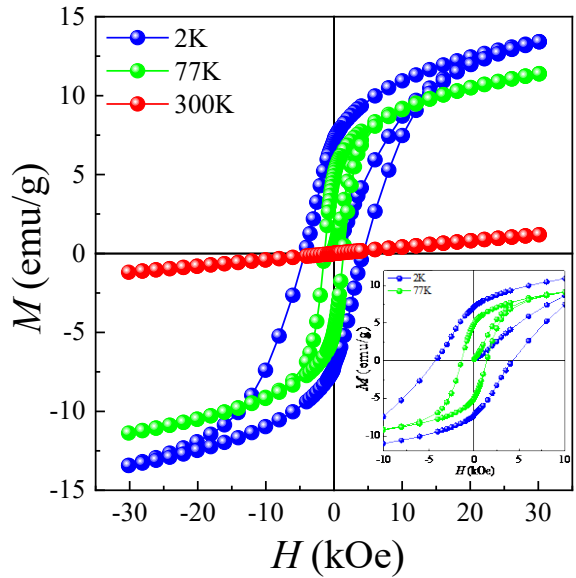

**Fig. 4.** Field dependences of magnetization $M(H)$ at 2, 77, and 300 K for the LSCO-900. The inset shows an enlarged area of the $M(H)$.

Table 3

**The main magnetic parameters: spontaneous magnetization $M_S$, residual magnetization $M_r$, coercivity $H_C$, characteristic Curie temperatures $T_{C1}$, $T_{C2}$, $T_{C3}$, $T_{C4}$, AFM transition temperature $T_t$, and blocking temperature $T_B$ under different pressures $P$ for the LSCO-900**

| P (GPa) | FM | | | | | | Near phase transitions | | | | | |
|---|---|---|---|---|---|---|---|---|---|---|---|---|
| | $M_S$ (emu/g) | | $M_r$ (emu/g) | | $H_C$ (kOe) | | FC | ZFC | FC | ZFC | ZFC | |
| | 2 (K) | 77 (K) | 2 (K) | 77 (K) | 2 (K) | 77 (K) | $T_{C1}$ (K) | $T_{C2}$ (K) | $T_{C3}$ (K) | $T_{C4}$ (K) | $T_{AFM}$ (K) | $T_B$ (K) |
| 0 | 10.5 | 8.7 | 7.2 | 4.9 | 4.3 | 1.4 | 251 | 255 | 163 | 187 | 145 | 169 |
| 0.16 | – | – | – | – | – | – | 259 | 261 | 161 | 186 | 150 | 164 |
| 0.43 | – | – | – | – | – | – | 260 | 263 | 162 | 190 | 153 | 168 |
| 0.64 | – | – | – | – | – | – | 265 | 267 | 165 | 191 | 156 | 170 |
| 0.76 | – | – | – | – | – | – | 266 | 269 | 166 | 194 | 158 | 170 |

The temperature dependences of magnetization $M_{ZFC}(T)$ and $M_{FC}(T)$ for the LSCO-900 under different pressures $P$ and in the magnetic field $H = 50$ Oe (Fig. 5) correspond to a typical behavior of the nanoparticles with FM and AFM inclusions, as well as with randomly oriented uniaxial magnetic anisotropy [63, 64]. From the temperature dependences of the extremum of the derivative of the magnetic susceptibility $d(M/H)/dT$, where $\chi = M/H$ (see the insets in Fig. 5), several characteristic phase transition temperatures were determined (Table 3). Noteworthy, four characteristic Curie temperatures $T_{C1} = 251$ K and $T_{C3} = 163$ K from FC and $T_{C2} = 255$ K and $T_{C4} = 187$ K from ZFC curves were defined as an example for the LSCO-900 without pressure $P = 0$ (see the insets in Fig. 5). The observed difference between ZFC and FC Curie temperatures is associated with "pinning" magnetic moments on the magnetic inhomogeneities and structural defects, as well as the competing different magnetic exchange interactions [18, 46, 61, 65]. The obtained average ZFC and FC Curie temperatures 175 and 253 K indicate the presence of smaller and bigger nanoparticles in the LSCO-900 [45, 66]. The highest Curie temperature, 253 K ($P = 0$), is compliant with other literature data 230 K [11] and 246 K [7] for the same ceramic bulk composition. An additional contribution to the ZFC and FC curves at $T < 187$ K is due to the magnetization from smaller particles. As shown [45, 66], a change in the Curie temperature $T_C$ value is caused by the particle size distribution. It means that in our case, the change of the average Curie temperatures in ~ 1.4 times corresponds to the existence of bigger $D_{XRD} \approx 60$ nm (see Table 2) with $T_C = 253$ K and smaller $D_{XRD} \approx 43$ nm with $T_C = 175$ K nanoparticles. Moreover, the blocking temperature $T_B = 169$ K, determined from the maximum temperature on the $M_{ZFC}(T)$ curve, is observed for the LSCO-900 ($P = 0$). At $T < T_B$, the magnetic nanoparticles are in a blocked state. At $T_B \leq T \leq T_{C2}$, the thermal fluctuations of the magnetization exceed the difference of the spin states between two energy minima, and the magnetic nanoparticles are in an unblocked state. At $T > T_{C2}$, the magnetic nanoparticles pass into the PM state. Additionally, an abnormal decreasing magnetization $M_{ZFC}(T)$ at an AFM transition temperature $T_{AFM}$ is observed (Fig. 5(b)) that is caused by AFM contribution of nanoparticles' "dead" skin [67].

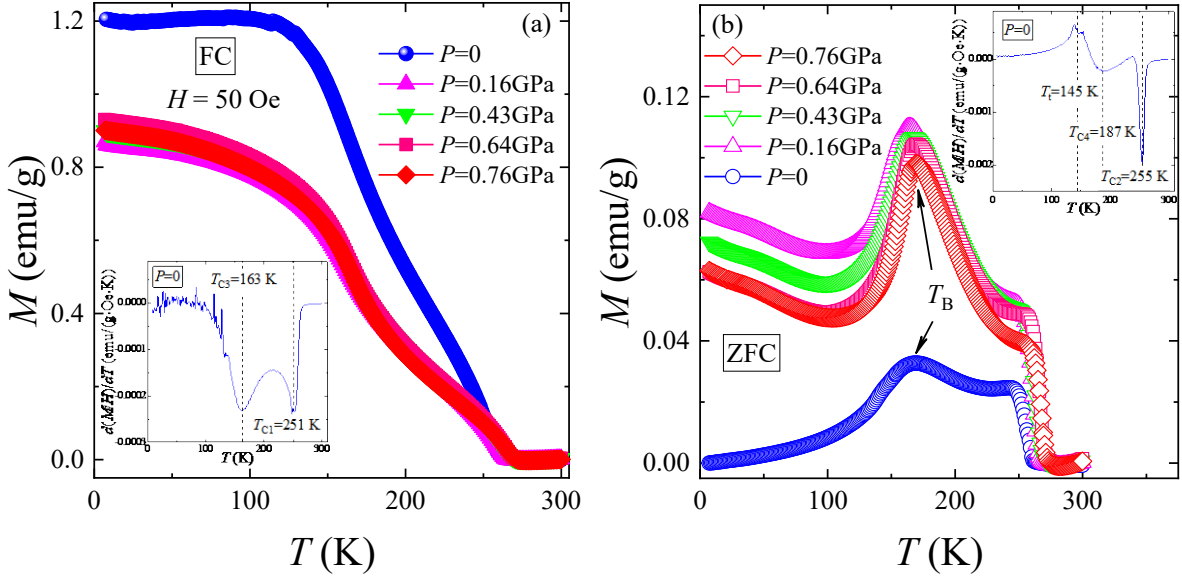

**Fig. 5.** Temperature dependences of the magnetization $M_{FC}(T)$ (a) and $M_{ZFC}(T)$ (b) for the LSCO-900 under different pressures $P$ and in the field of $H = 50$ Oe. The insets show the defined characteristic Curie temperatures $T_{C1}$ and $T_{C3}$ (a) and $T_{C2}$ and $T_{C4}$ (b), as well as AFM transition temperature $T_{AFM}$ and blocking temperature $T_B$ (b).

As pressure $P$ increases, the FM and AFM subsystems become more magnetically homogeneous. The characteristic Curie temperatures $T_{C1}$ and $T_{C2}$ for bigger and $T_{C3}$ and $T_{C4}$ for smaller nanoparticles increase with a pressure rise (Table 3), indicating the strengthening of the FM DE $Co^{3+}$–$O^{2-}$–$Co^{4+}$. At the same time, the $T_{AFM}$ also increases, demonstrating the improvement of the AFM superexchanges $Co^{3+}$–$O^{2-}$–$Co^{3+}$ and $Co^{4+}$–$O^{2-}$–$Co^{4+}$, whereas the blocking temperature $T_B$ increases slightly (Table 3). The characteristic Curie temperatures show a linear-like behavior *versus* pressure (Fig. 6). This allows us to define an average $<dT_C/dP>$ ratio for smaller and bigger nanoparticles being 10 and 13 K/GPa, respectively. The obtained values show that bigger nanoparticles are more sensitive to the external pressure $P$, and their magnetic parameters can be easily controlled.

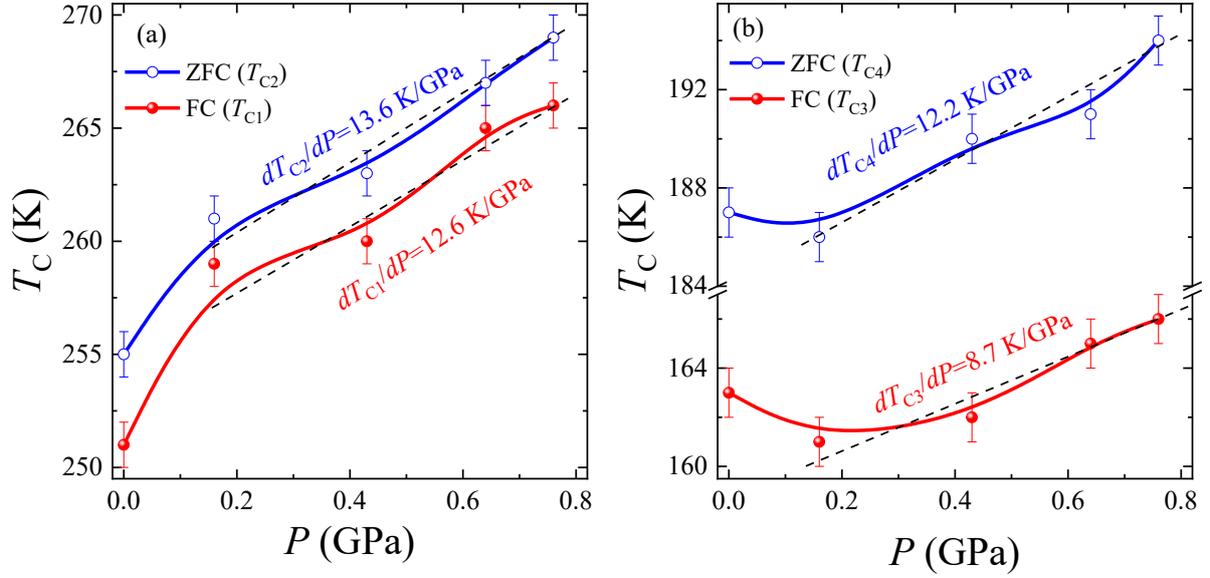

**Fig. 6.** Pressure dependences of characteristic Curie temperatures $T_{C1}$ and $T_{C2}$ (a) and $T_{C3}$ and $T_{C4}$ (b) defined from the ZFC-FC curves for the LSCO-900. The dashed lines show linear behavior $T_C$ versus $P$.

The isothermal magnetization curves $M(H)$ around $T_C$ in a wide temperature range from 200 to 300 K with a $\Delta T = 2$ K step for the LSCO-900 are shown in Fig. 7(a). At $T > T_C$, the magnetization increases linearly with an applied magnetic field, indicating the PM state of the sample. At $T < T_C$, the linear isothermal magnetization behavior turns to a curve, which indicates passing to the FM state. At the same time, the magnetization cannot be saturated even at lower temperatures and higher magnetic fields, probably because of AFM phase contribution [68]. According to the Banerjee criterion, [69] the positive slope of Arrott's curves $M^2(H/M)$ (see Fig. 7(b)) near the Curie temperature is observed, which indicates a second-order magnetic phase transition for the LSCO-900.

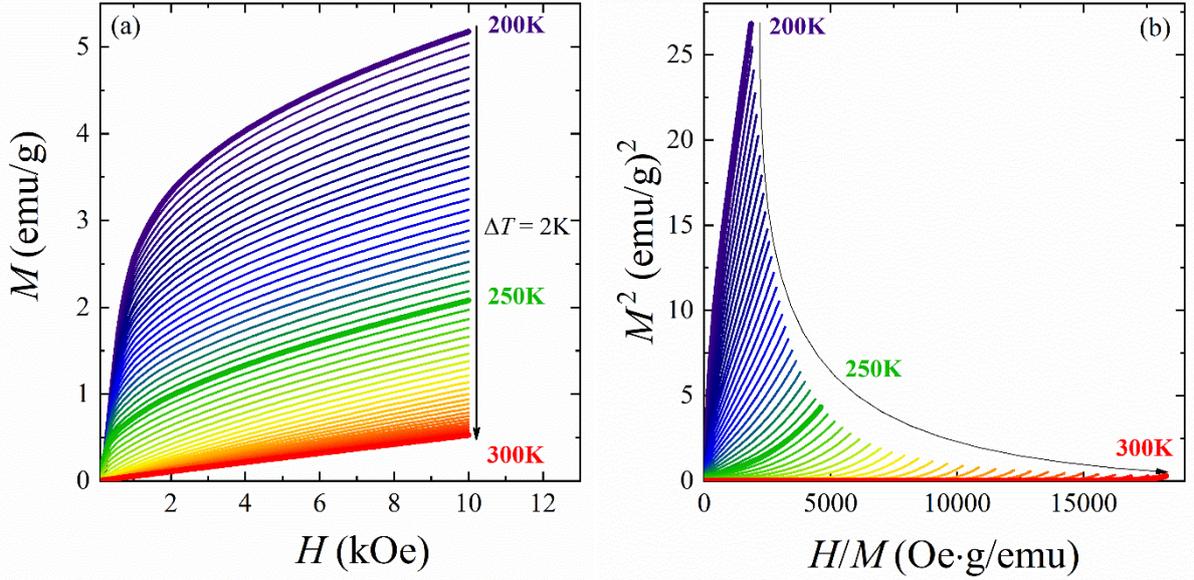

**Fig. 7.** Isotherms of magnetization $M(H)$ (a) and Arrott's plots $M^2(H/M)$ (b) within 200–300 K with a $\Delta T = 2$ K step for the LSCO-900.

The temperature dependences of the magnetic entropy change $-\Delta S_M(T, H)$ near phase transition(s) (Fig. 8) were plotted using the isotherms of magnetization $M(H)$ and the numerical integration method of the Maxwell relation [70]:

$$\Delta S_M(T, \mu_0 H) = S_M(T, \mu_0 H) - S_M(T, 0) = \int_0^{\mu_0 H} (\partial M / \partial T)_{\mu_0 H} d(\mu_0 H). \tag{1}$$

The magnetic entropy change $-\Delta S_M(T, H)$ shows a vast indistinct peak with relatively low values of MCE. Upon application of the external high hydrostatic pressure $P \approx 0.8$ GPa and with increasing the magnetic field $\Delta H$ up to 10 kOe, the MCE peak position moves towards higher temperatures from 250 to 256 K, indicating the domination of the FM DE over AFM superexchange interactions, whereas without pressure $P = 0$ we have opposite situation and reducing from 247 to 233 K (Fig. 8). Moreover, the $\delta T_{FWHM}$ for the $-\Delta S_M^{max}$ is far more than 50 K that makes the studied LSCO-900 is a promising component for creating the magnetic refrigerant composites working a wide temperature range.

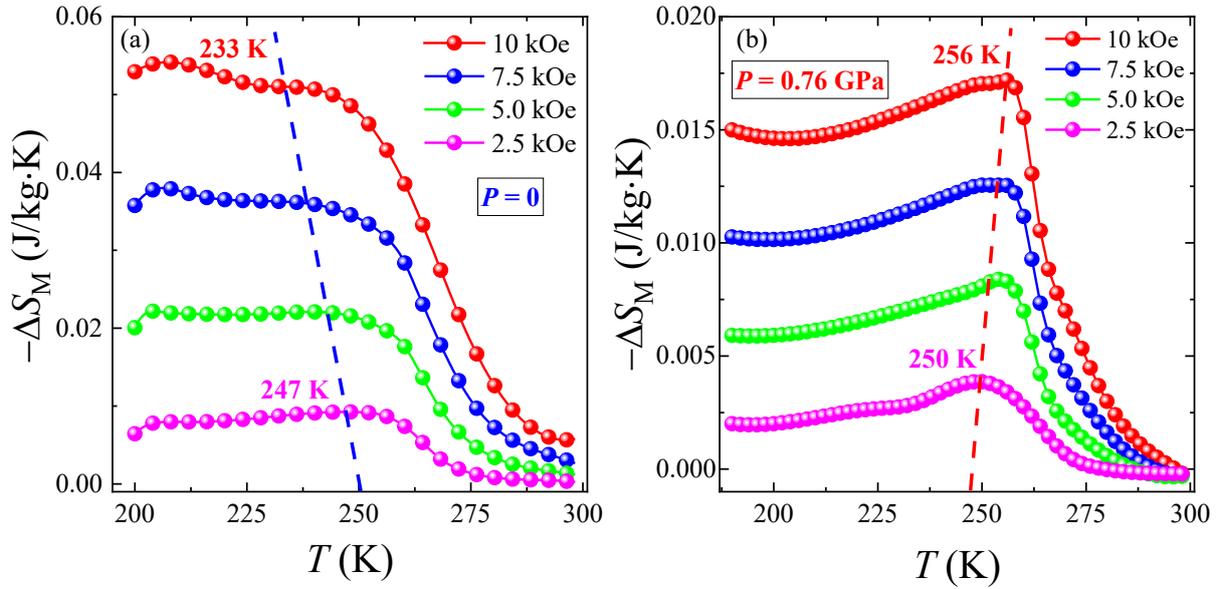

**Fig. 8.** Temperature dependences of magnetic entropy change $-\Delta S_M(T)$ without pressure $P = 0$ (a) and under high-pressure $P \approx 0.8$ GPa (b) in the magnetic field $\Delta H$ up to 10 kOe for the LSCO-900.

### 3.3. Electrocatalytic properties

As mentioned in the introduction, interest in Ln- and Co-containing perovskites as electrocatalysts for OER in an alkaline environment has increased significantly recently [28, 30-36, 41-44]. In order to find out the OER activity for water splitting of the prepared LSCO samples, electrochemical tests were carried out in 1M KOH electrolyte (pH = 14) with the following sequence: multiple current-voltage (CV) scans (until stable electrode operation is achieved), alternative current (AC) impedance, recording of LSV curves with iR correction, long-term chronopotentiometry tests (at a current density of 10 mA·cm$^{-2}$) and re-registration of LSV curves. It should be emphasized that Co- and Ni-containing complex oxides can be entirely or partially *in situ* transformed into oxides/hydroxides due to electrochemical processes in an alkaline medium [71, 72]. Accordingly, after a long electrochemical treatment, we investigated the electrode materials to verify the stability of the LSCO catalysts.

Before the electrochemical test, the manufactured electrodes were placed in an electrolyte environment for 1 h to establish equilibrium at the electrode-electrolyte interface. As shown in Fig. S2, for all samples, the shape of the CV curves in the range of 0.9–1.8 V is similar at the corresponding scanning rate (from 1 to 50 mV·s$^{-1}$). Only the peak around 1.3 V (anodic scans) is clearly observed, corresponding to the $Co^{4+} \rightarrow Co^{3+}$ transformation [47]. As AC impedance measurements show (Fig. 9(a)), the LSCO-900 sample has the best conductivity. Nyquist plots correspond to the equivalent circuit, which includes series resistance $R_s$, the EDLC at the active material/electrolyte interface $C_{dl}$, electron transfer resistance $R_{ct}$, and the diffusion impedance Warburg impedance (Fig. S3). The impedance spectra include two parts, a line in a low-frequency region and a semicircle in a high-frequency region. The numerical value of the diameter of the semicircle on the $Z_0$ axis is approximately equal to the charge transfer resistance $R_{ct}$, and the sloping line corresponds to the Warburg impedance, which is associated with electrolyte diffusion in electrode materials. As known, the Helmholtz double layer capacitance $C_{dl}$ appears due to the porosity of the active material and the uneven distribution of the current on the electrode surface [73]. The double-layer capacitance $C_{dl}$ was measured by cyclic voltammetry for determined the electrochemically active surface area (ECSA) (see details in ESI2). Fig. S4 shows CV curves recorded in the potential windows from 0 to 0.1 V vs. Hg/HgO at a scan rate of 1–10 mV·s$^{-1}$. It is known that the ECSA value corresponds to the number of potential active sites for electrocatalysis, and $R_{ct}$ is inversely proportional to electrochemical activity. In this way, the calculated values ESCA for the LSCO-800, LSCO-850, and LSCO-900 were 15.97 cm$^2$, 17.26 cm$^2$, and 24.53 cm$^2$, respectively, whereas the resistance values $R_{ct}$ were 1.05, 0.9, and 0.6 Ohm. Thereby, a low charge transfer resistance and high ESCA indicate good electronic conductivity and a high rate of OH ion transfer across the boundaries between the electrolyte and active electrode materials. The above AC impedance results show that the LSCO-900 has the attributes of an efficient electrocatalyst for OER in an alkaline environment.

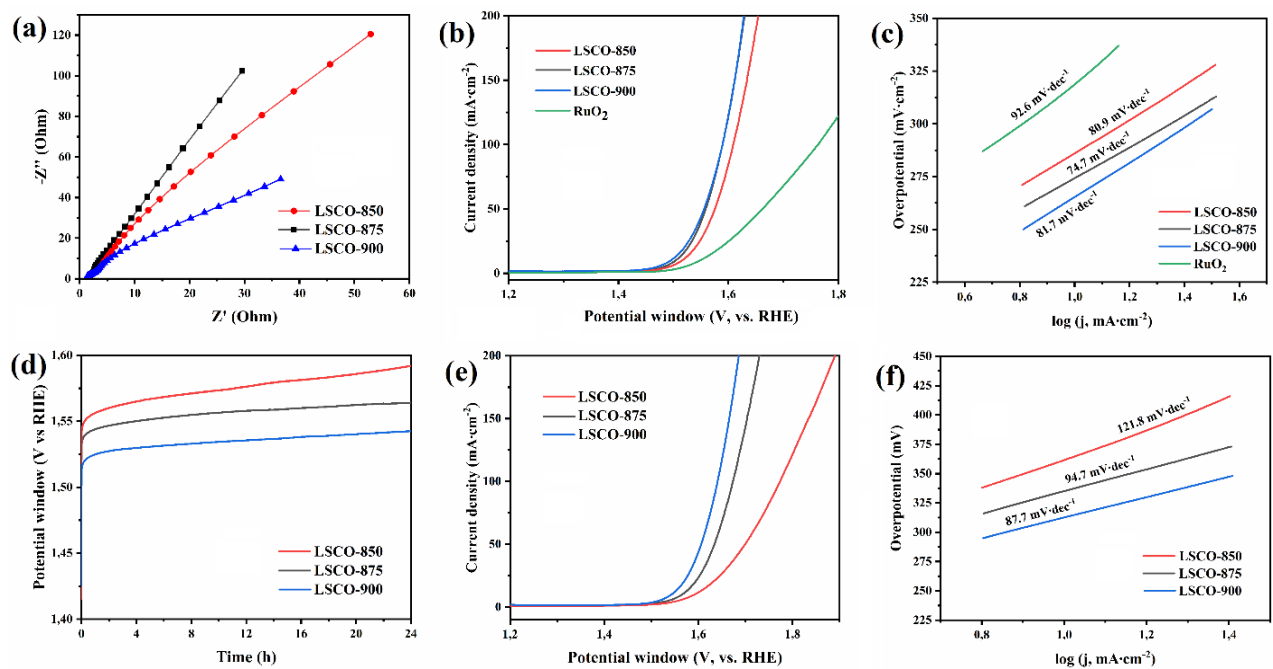

**Fig. 9**. Electrochemical testing for the LSCO electrodes in 1 M KOH: (a) Nyquist plots; (b) and (c) LSV curves and Tafel plots before the long-term stability test for the LSCO catalysts and $RuO_2$ at a scan rate of 1 mV s$^{-1}$ (iR-compensation 90%); (d) time dependence of potential under a constant current density of 10 mA cm$^{-2}$; (e) and (f) LSV curves and Tafel plots after the long-term stability test at a scan rate of 1 mV s$^{-1}$.

According to onset LSV measurements, all LSCO samples exhibited highly efficient electrocatalytic activity for alkaline OER and a slow boost of current density with the increase of applied potential (Fig. 9(b)). The obtained overpotentials values' intervals were only 265–285 mV and 360–380 mV when reaching a current density of 10 and 100 mA cm$^{-2}$, respectively (more details are given in Table S2). The LSV curves for the LSCO-900 and LSCO-875 electrodes are almost identical, and the overpotential OER for the LSCO-850 is slightly higher. It is important that the commercial electrocatalyst of noble metal oxide ($RuO_2$) used for comparison showed worse activity. These results confirm the calculations of the Tafel slope (Fig. 9(c)): for the synthesized LSCO samples, the range of values is 74–82 mV·dec$^{-1}$, and for commercial $RuO_2$ is slightly more than 92 mV·dec$^{-1}$. The subsequent long-term test of electrodes during continuous electrolysis (current density 10 mA·cm$^{-2}$) is presented in Fig. 9(d). The continuous formation of oxygen bubbles on the surface of the electrodes was observed when the stability tests were carried out. For all samples, an increase in overpotential OER is observed during the first 4 h of the chronopotentiometry test, and the values of $\eta_{10}$ reach 300, 320, and 335 mV for the LSCO-900, LSCO-875, and LSCO-850, respectively. However, the activity of the electrocatalysts decays much more slowly during the next 20 h of electrolysis, especially in the case of the LSCO-900 sample (the magnitude of the overpotential increases by only 12 mV); more detailed these results for all electrodes presented in Table S3. The slight increase in voltage after 4 h of the test proves the final stabilization of all electrodes and the high efficiency of the LSCO materials as OER catalysts. It should also be considered that carbon fiber (the base of the electrode used in this study) is gradually destroyed in an alkaline environment under OER conditions [72]. This leads to breakdowns of the interface between the catalyst and the collector-electrode and is an additional factor in the growth of overpotential OER during long-term tests. LSV curves measured after long-term stability tests (Fig. 9(e)) are in good agreement with CP results: the values of overpotential $\eta_{10}$ for the corresponding samples are identical (Tables S2 and S3). Re-registration LSV measurements revealed that in the range of current densities from 10 to 100, the OER overvoltage increased by only 14–17% for the LSCO-900 electrode, while for the LSCO-875 and LSCO-850, the changes were 22–24 and 27–44%, respectively (Table S2). Calculation of the Tafel slope after long-term electrolysis (Fig. 9(f)) also indicates the preservation of high catalytic activity of the LSCO-900 material, in contrast to the LSCO-875 and especially LSCO-850 samples.

Electrode materials were also studied in detail after CP tests. According to the XRD, all LSCO samples retained their crystallinity. As shown in Fig. S5, the diffraction patterns show reflections from both crystalline phases ($La_{0.6}Sr_{0.4}CoO_3$ and $SrCo_{0.78}O_{2.48}$). This proves the high stability of Ln and Co-containing complex oxides in an alkaline environment during long-term electrolysis, in contrast to similar Mn-containing compounds [45]. SEM and elemental mapping of the electrode surface after prolonged electrolysis mostly support this conclusion (Fig. S6–S8). The SEM images mostly show nanoparticles as in the initial LSCO material, which agrees with the elemental mapping. However, a careful analysis of the images revealed that the formation of nanoflakes is observed in some areas (Fig. S7). In addition, the XPS study was performed before and after the electrochemical tests to elucidate the features of possible transformations on the surface of the LSCO for the electrode with the best catalytic performance (LSCO-900). In both cases, the survey spectra of the LSCO-900 electrode show the existence of La, Sr, Co, O, and C peaks (Fig. S9). As expected, La and Sr ions are trivalent and divalent before and after electrocatalysis (Fig. S9 and Table S4).

Nevertheless, the high-resolution XPS spectra revealed some differences (a detailed description is provided in the ESI3). Changes are also observed for the HR-XPS spectra of Co2p (Fig. 10): exhibit two prominent peaks at around 780.1 and 795.6 eV for the electrode before and at around 778.8 and 794 eV after electrocatalysis, which correspond to the $Co2p_{3/2}$ and $Co2p_{1/2}$ states, respectively. Both Co2p spectra were decomposed into two components (Fig. 10(b, c)), suggesting cobalt ions' presence in $Co^{3+}$ and $Co^{4+}$ oxidation states. The positions of their peak centers are listed in Table S5 (ESI3) and are in good agreement with other literature XPS data [16, 74, 75]. The determined values of the $Co^{3+}/Co^{4+}$ ratio differ before 52.4/47.6 and after 61.8/38.2 electrocatalysis. It indicates the growth of the $Co^{3+}$ content and is only a quality picture of the cobalt ions content due to the presence of additional weak $Co^{2+}$ at around 786 eV and $Co_3O_4$ at around 790 eV satellite peaks [76]. In support of the $Co^{3+}$ content growth, the following factors should be noted. First, the LSCO-900 spectrum after electrocatalysis is slightly shifted towards the lower energy ~ 1.3 eV relative to the one before electrocatalysis, indicating an increase of the $Co^{3+}$ concentration. Second, the simultaneous reducing $Co^{2+}$ satellite peak and, as a result, its content and increasing $Co_3O_4$ satellite peak, consisting of the mixed $Co^{2+}$ and $Co^{3+}$ states, is observed for the sample after electrocatalysis. Third, the full-width at half-maximum (FWHM) of the $Co2p_{3/2}$ main peak decreases ~ 2.0 eV after electrocatalysis, suggesting more $Co^{3+}$ and fewer $Co^{2+}$ ions on the surface [77]. Therefore, the XPS study clearly indicates significant changes in the surface layer of the LSCO catalyst after long-term electrolysis in an alkaline environment. Because the crystallinity and morphology of the original LSCO material remain intact, the formation of a dense amorphized layer with high catalytic activity on the surface of the LSCO should be assumed, essentially the formation of a core-shell structure.

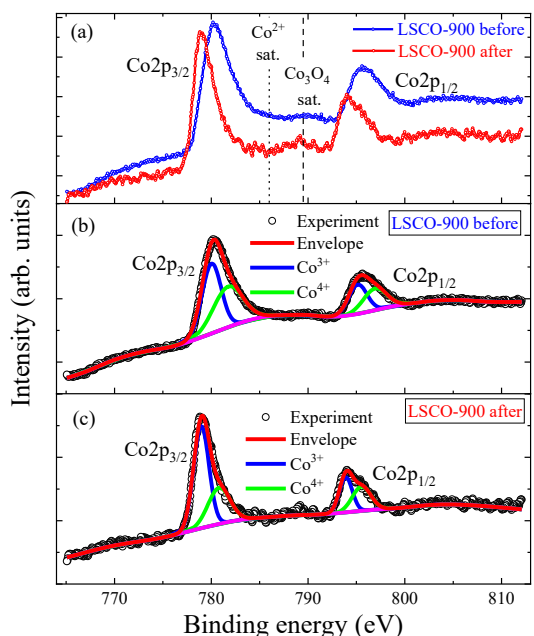

**Fig. 10**. Co2p spectra (cycles) for the LSCO-900 before and after electrocatalysis (a), as well as their fitting by red envelope line and decomposition into $Co^{3+}$ (blue line) and $Co^{4+}$ (green line) components with a baseline (magenta line) in the LSCO-900 before (b) and after (c) electrocatalysis. The dashed and dotted lines indicate the energy position of the $Co_3O_4$ and $Co^{2+}$ satellite peaks, respectively.

Table 4 lists the comparative characteristics of OER for electrocatalysts based on La, Co-containing perovskites in an alkaline electrolyte. Among them, the LSCO material described here occupies a leading position. It should be noted that very recently, it was reported that the amorphization of crystallized nanometer perovskite $LaCoO_3$ to a highly disordered state leads to improved electrocatalytic properties of OER [44]. In our case, a highly active amorphized layer is formed on the surface of the LSCO nanoparticles during electrolysis. The catalytic activity increases with increasing the proportion of this crystalline phase for a number of the LSCO-850, LSCO-875, and LSCO-900 samples. In addition, the high stability of the amorphized layer prevents further destruction of the nanoparticles and protection of the initial morphology, which can provide a significant electroactive surface area for a long time. Finally, forming *in situ* core-shell nanostructures based on perovskites is a potentially successful strategy for creating highly active and stable electrocatalysts.

Table 4
Comparison of OER properties for LaCo-based electrocatalysts in alkaline electrolyte.

| Catalyst | Electrolyte | $\eta_{10}$ (mV) | Tafel slope | Ref. |
|---|---|---|---|---|
| LSCO-900 | 1M KOH | 265<br>312* | 81.7<br>87.7* | This work |
| LSCO-850 | 1M KOH | 275<br>335* | 74.7<br>94.7* | This work |
| LSCO-875 | 1M KOH | 285<br>362* | 80.9<br>121.8* | This work |
| $LaCoO_3$-Reduce | 1M KOH | 293 | 63.4 | [44] |
| $La_{0.5}Pr_{0.5}CoO_3$ | 1M KOH | 312 | 80.6 | [28] |
| $La(Co_{0.71}Ni_{0.25})_{0.96}O_3$ | 0.1M KOH | 324 | 71 | [29] |
| $La(CrMnFeCo_2Ni)O_3$ | 1M KOH | 325 | 51.2 | [30] |
| $La_{0.7}Sr_{0.3}CoO_{3-P}$ | 1M KOH | 326 | 70.8 | [31] |
| Cl-$LaCoO_3$ | 1M KOH | 342 | 76.2 | [42] |
| $(La_{0.776}Sr_{0.224})_{0.9}(Co_{0.087}Fe_{0.84}Ru_{0.073})_{1.026}O_{3-\delta}$ | 1M KOH | 347 | 54.7 | [32] |
| La-$CoO_x$ | 0.1M KOH | 353 | 78.2 | [41] |
| $La_{0.4}Sr_{0.6}Co_{0.7}Fe_{0.2}Nb_{0.1}O_{3-\delta}$ | 0.1M KOH | 360 | 78 | [33] |
| $La_{0.9}CoO_{3-\delta}$ | 0.1M KOH | 380 | 82.5 | [78] |
| $La_{0.96}Ce_{0.04}CoO_3$ | 1M KOH | 380 | 80 | [34] |
| $La_{0.6}Sr_{0.4}Co_{0.8}Fe_{0.2}O_3$ | 0.1M KOH | 385 | 76.7 | [35] |
| $LaCoO_3$ | 0.5M KOH | 396 | 144.8 | [79] |
| Pt/$LaCoO_3$ | 1M KOH | 327 | 92 | [43] |
| $LaMn_{0.4}Co_{0.6}O_3$ | 1M KOH | 400 | 95 | [36] |
| $La_{0.699}Sr_{0.301}Co_{0.702}Fe_{0.298}O_{2.92}$ | 1M KOH | 440 | 109 | [37] |
| $LaCoO_3$ | 0.1M KOH | 490 | 69 | [39] |

*After 24 h of incessant electrolysis (current density 10 mA·cm$^{-2}$).

## 4. Conclusions

The structure, morphology, particle size distribution, and magnetic phase transitions, as well as magnetic, magnetocaloric, and electrochemical properties of the LSCO nanopowders obtained under different annealing temperatures $t_{ann}$ = 850, 875, and 900 °C have been studied comprehensively. All samples indicate the rhombohedral $R\bar{3}c$ perovskite structure, improving its single-phase nature with increasing $t_{ann}$. At the same time, it leads to increasing unit cell volume $V$, an average particle size $D$, and its dispersion $\sigma$. The LSCO-900 composition mainly consists of the $Co^{3+}$ and $Co^{4+}$ ions, as well as minor $Co^{2+}$ and $Co_3O_4$ traces that play a crucial role in forming magnetic and electrochemical properties. The main magnetic parameters such as spontaneous magnetization $M_S$, coercivity $H_C$, residual magnetization $M_r$, phase transition temperatures, and magnetic entropy change $-\Delta S_M$ have been defined. The LSCO-900 sample is a ferromagnet with a high $H_C$ = 4.3 kOe. It has been found that the LSCO-900 exhibits two characteristic Curie temperatures, $T_{C1}$ and $T_{C2}$, which are associated with the presence of nanoparticles of different sizes. Additionally, FM and AFM subsystems are stabilized and become more magnetically homogeneous because of applied external hydrostatic high-pressure $P$. As pressure increases, the average $<T_{C1}>$ and $<T_{C2}>$ increase from 253 and 175 K under ambient pressure to 268 and 180 K under $P \approx 0.8$ GPa, respectively. At the same time, the AFM transition temperature $T_{AFM}$ and blocking temperature $T_B$ also increase from 145 and 169 K to 158 and 170 K, respectively. The rise of $<dT_{C1}/dP>$ for bigger particles and $<dT_{C2}/dP>$ for smaller particles differs and is 13 and 10 K/GPa, respectively. The MCE in the LSCO-900 nanopowder is near the Curie temperature and has a second-order phase transition. The values of $-\Delta S_M$ are quite low and equals 0.05 J/kg·K without pressure and 0.017 J/kg·K under $P \approx 0.8$ GPa in the magnetic field 1 T, but with extremely wide peak $\delta T_{FWHM} > 50$ K.

The obtained series of the LSCO nanomaterials showed excellent characteristics as electrocatalysts for overall water splitting (OER process) in 1 M KOH electrolyte. The values of the initial overpotential of oxygen generation were within 265–285 mV (a current density of 10 mA/cm$^2$), and the sample with the highest content of the La$_{0.6}$Sr$_{0.4}$CoO$_3$ phase showed the best performance. The LSCO materials described here retain high catalytic activity during long-term electrolysis, which should be attributed to the formation of a stable amorphization layer on the surface of nanoparticles. Accordingly, a significant area of the catalytically active surface, resulting from the emergence of stable core-shell nanosystems, can ensure high efficiency of practical use of the material.

The parallel study of nanopowders' physical and electrocatalytic properties in this work made it possible to establish the physical and chemical basis of the various samples obtained at different $t_{ann}$ and find ways to obtain materials for practical application in electrocatalysis. The research of magnetic and magnetocaloric properties of the samples opens up the possibility of changing the conditions of catalysis by changing the temperature of the electrocatalyst by a contactless method using an electromagnetic field. The obtained results demonstrate a simple way to improve complex oxides with a perovskite structure for various functional purposes.

## Author Contributions

Han Lin: Writing - original draft, Data curation, Investigation, Methodology. N.A. Liedienov: Writing - review & editing, Conceptualization, Data curation, Investigation. I.V. Zatovsky: Writing - review & editing, Data curation, Conceptualization, Investigation. D.S. Butenko: Writing - review & editing, Data curation, Investigation, Methodology. I.V. Fesych: Data curation, Investigation, Methodology. Wei Xu: Data curation, Investigation, Methodology. Songchun Rui: Data curation, Investigation, Methodology. Quanjun Li: Data curation, Investigation, Methodology. Bingbing Liu: Data curation, Investigation, Methodology. A.V. Pashchenko: Data curation, Methodology. G.G. Levchenko: Writing - review & editing, Supervision.

## Conflicts of interest

There are no conflicts to declare.

## Acknowledgments

This research was financially sponsored by the European Federation of Academies of Sciences and Humanities within the framework of the "European Fund for Displaced Scientists" (Grant reference number EFDS-FL2-05). It was also supported by the Major Science and Technology Infrastructure Project of Material Genome Big-science Facilities Platform supported by the Municipal Development and Reform Commission of Shenzhen.

**The multifunctionality of lanthanum-strontium cobaltite nanopowder: high-pressure magnetic and excellent electrocatalytic properties for OER**

Hanlin Yu[a], N.A. Liedienov[a,b,*], I.V. Zatovsky[c], D.S. Butenko[d,e,*], I.V. Fesych[f], Wei Xu[g], Songchun Rui[h], Quanjun Li[a], Bingbing Liu[a], A.V. Pashchenko[a,c,i], G.G. Levchenko[a,c,*]

[a]State Key Laboratory of Superhard Materials, International Center of Future Science, Jilin University, 130012 Changchun, P.R. China
[b]Donetsk Institute for Physics and Engineering named after O.O. Galkin, NASU, 03028 Kyiv, Ukraine
[c]F.D. Ovcharenko Institute of Biocolloidal Chemistry, NASU, 03142 Kyiv, Ukraine
[d]Shenzhen Key Laboratory of Solid State Batteries, Southern University of Science and Technology, Shenzhen 518055, P.R. China
[e]Academy for Advanced Interdisciplinary Studies, Southern University of Science and Technology, Shenzhen 518055, P.R. China
[f]Taras Shevchenko National University of Kyiv, 01030 Kyiv, Ukraine
[g]State Key Laboratory of Inorganic Synthesis and Preparative Chemistry, College of Chemistry, Jilin University, Changchun, 130012, P.R. China
[h]Baicheng Normal University, 137099 Baicheng, China
[i]Institute of Magnetism NASU and MESU, 03142 Kyiv, Ukraine

*Corresponding author
E-mail address:  nikita.ledenev.ssp@gmail.com (N.A. Liedienov)
 debut98@ukr.net (D.S. Butenko)
 g-levch@ukr.net (G.G. Levchenko)

**ESI1**
**Determination of the size of the coherent scattering region in the La$_{0.6}$Sr$_{0.4}$CoO$_3$ nanopowders**

The size of the coherent scattering region $D_{XRD}$ was determined using the X-ray line broadening method. The average size $D_{XRD}$ in the La$_{0.6}$Sr$_{0.4}$CoO$_3$ (LSCO) nanopowders is related to the dimensional broadening of $\beta$ for diffraction reflection (012) according to the Scherrer equation [1]:

$$D_{XRD} = K\lambda / \beta\cos\theta, \tag{S1}$$

where $D_{XRD}$ is the size of scattering crystallites in nm; $\lambda = 0.15406$ nm is the wavelength of X-ray radiation; $K = 0.9$ is a constant that depends on the method for determining the line broadening and crystal shape; $\beta$ is the width of the intensity distribution curve at half of the height of the maximum of the reflex in radians; $\theta$ is the diffraction angle in degrees.

Considering that the integral width of the peak in the diffractogram is approximated by the pseudo-Voigt function with a large (up to 90% or more) contribution of the Lorentz function, the Lorentzian was used to describe the shape of the diffraction reflection at $2\theta \approx 23.2°$ (see Fig.S1). In order to exclude the instrumental broadening $\beta_{inst}$, standard silicon (Si) X-ray powder diffraction data (JCPDS89-2955) was recorded under the same condition in a separate experiment. The integral width of the peak was calculated by the formula [2]:

$$\beta_{012} = \beta_{exp} - \beta_{inst}, \tag{S2}$$

where $\beta_{exp}$ is the experimental width of the sample peak at half the maximum intensity; $\beta_{inst}$ is an instrumental broadening of the diffraction line, which depends on the design features of the diffractometer (in radians).

The average size of the $D_{XRD}$ was obtained using an approximation of the experimental values of the intensity of the diffraction maximum with a Bragg angle of $2\theta \approx 23.2°$ and considering all the experimental parameters in equation (S1) (see Table S1).

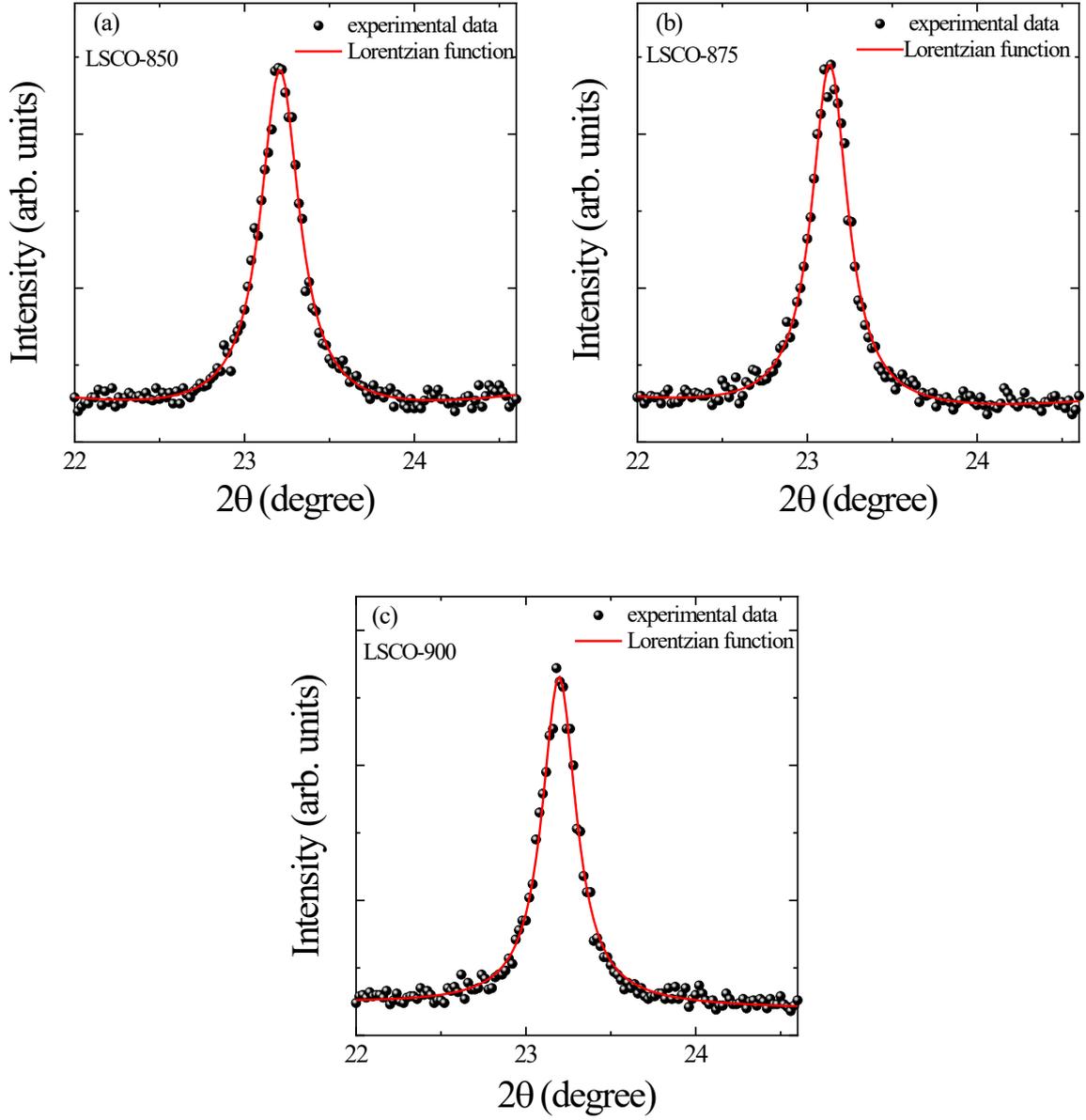

**Fig. S1**. The diffraction patterns and their approximation by Lorentzian function for the LSCO nanopowders in the region of the (012) reflection with an angle of $2\theta \approx 23.2°$.

**Table S1**
The experimental parameters in Eq. (S1) and the average size of the coherent scattering regions $D_{XRD}$ for the LSCO nanopowders with different annealing temperatures $t_{ann}$.

| $t_{ann}$ (°C) | $2\theta$ (degree) | $\beta$ (radian) | $\cos\theta$ | $\lambda$ (nm) | $K$ | $D_{XRD}$ (nm) |
|---|---|---|---|---|---|---|
| 850 | 23.208 | 0.0030 | 0.97956 | 0.15406 | 0.9 | 48±1 |
| 875 | 23.132 | 0.0026 | 0.97969 | 0.15406 | 0.9 | 54±1 |
| 900 | 23.195 | 0.0024 | 0.97958 | 0.15406 | 0.9 | 60±1 |

## Electrochemical properties of the LSCO nanopowders

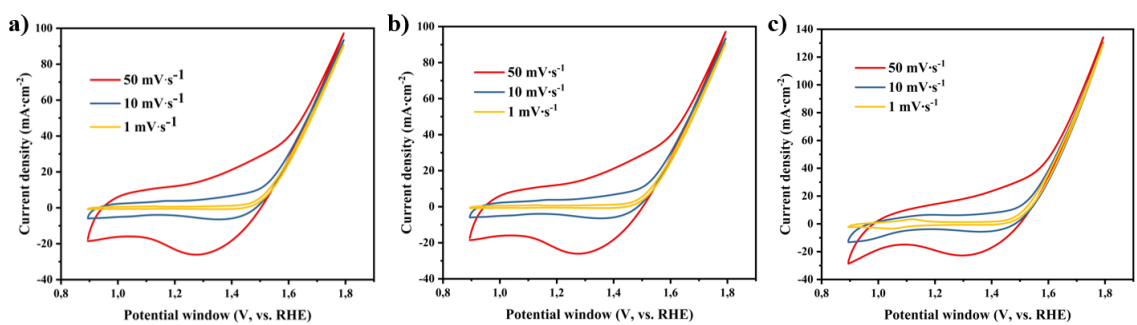

**Fig. S2**. CV curves at different scanning rates for LSCO electrodes: (a) LSCO-850, (b) LSCO-875, (c) LSCO-900.

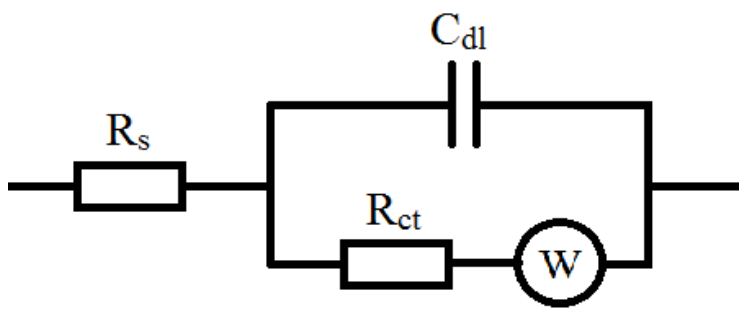

**Fig. S3**. The equivalent scheme for EIS measurement.

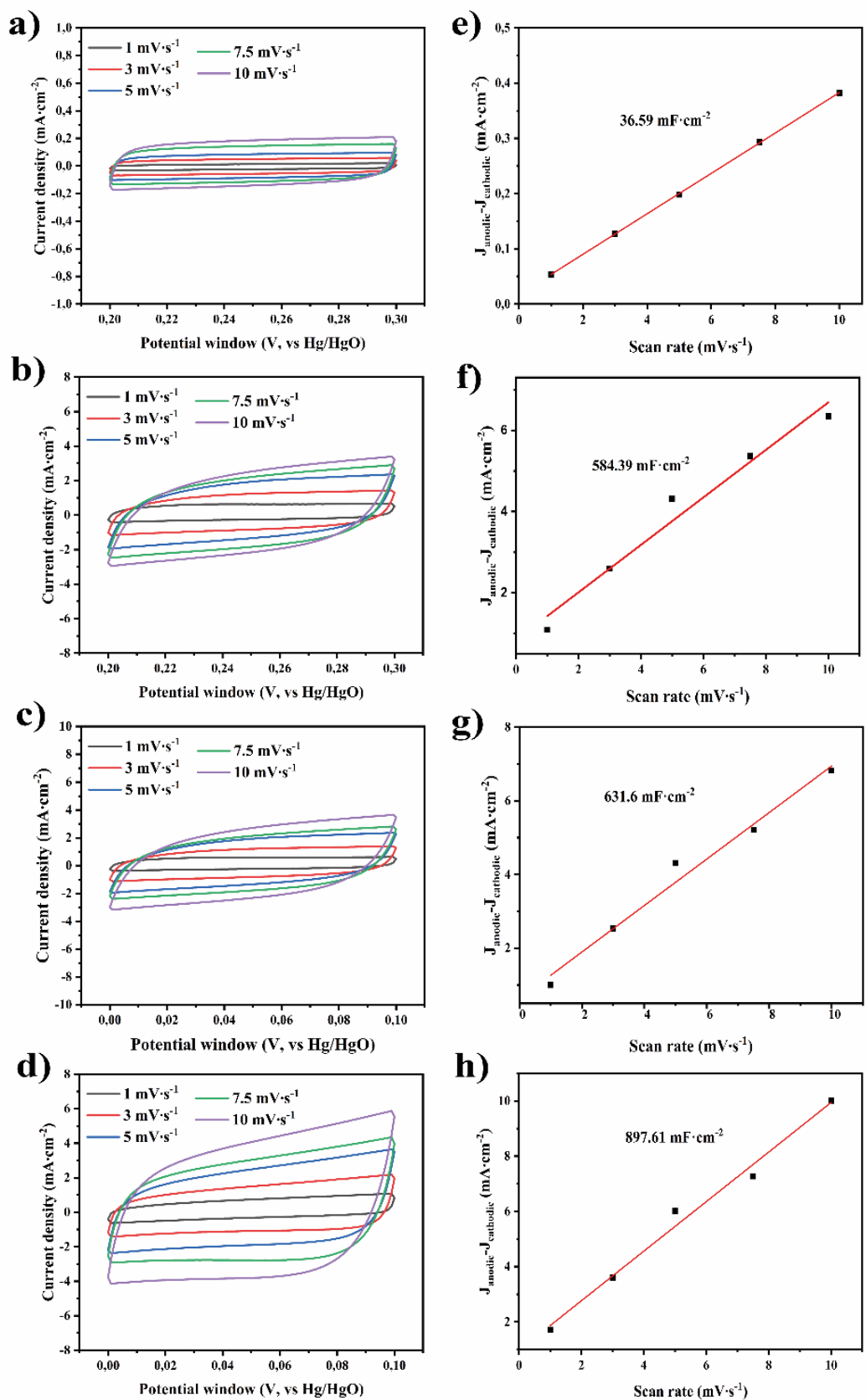

**Fig. S4**. Cyclic voltammograms of (a) LSCO-850, (b) LSCO-875, (c) LSCO-900, and (d) carbon fiber at different scan rates in the voltage range of 0–0.1 V. Δj *versus* RHE as a function of the scan rate for the (e) LSCO-850, (f) LSCO-875, (g) LSCO-900, and (h) carbon fiber.

The double-layer capacitance $C_{dl}$ was estimated by plotting the $\Delta j$ *versus* RHE as a function of the scan rate (Fig. S4 (e-h)): $C_{dl} = d(\Delta j)/(2dV)$. The ECSA can be calculated from the $C_{dl}$ according to ECSA = $C_{dl}/C_s$ ($C_s$ is the specific capacitance of a flat surface with 1 cm² of real surface area). In this case, CC-based with Super P electrode can be accepted as a standard, showing the measured capacitance of 36.59 mF·cm⁻². The ECSA values of LSCO-850, LSCO-875, and LSCO-900 are 15.97 cm², 17.26 cm², and 24.53 cm², respectively.

**Table S2**
**Overpotential OER and Tafel slope values for LSCO samples according to LSV tests.**

| Sample | Overpotential OER | | | Tafel slope (mV·dec⁻¹) |
|---|---|---|---|---|
| | $\eta_{10}$ (mV) | $\eta_{20}$ (mV) | $\eta_{100}$ (mV) | |
| *Initial values* | | | | |
| LSCO-850 | 285 | 310 | 379 | 74.7 |
| LSCO-875 | 275 | 296 | 360 | 80.9 |
| LSCO-900 | 265 | 290 | 360 | 81.7 |
| RuO₂ | 320 | 356 | 534 | 92.6 |
| *After 24 h of incessant electrolysis (current density 10 mA·cm⁻²)* | | | | |
| LSCO-850 | 362 | 401 | 543 | 121.8 |
| LSCO-875 | 335 | 363 | 444 | 94.7 |
| LSCO-900 | 312 | 338 | 410 | 87.7 |
| *Change in LSV value after CP test (current density 10 mA·cm⁻²), %* | | | | |
| LSCO-850 | +27.0 | +29.4 | +43.3 | – |
| LSCO-875 | +21.8 | +22.6 | +23.3 | – |
| LSCO-900 | +17.7 | +16.6 | +13.9 | – |

**Table S3**
**The value of overpotential OER during electrolysis for LSCO samples according to the results of the CP tests.**

| Sample | Electrolysis time (h) | | | | |
|---|---|---|---|---|---|
| | 0 | 1 | 4 | 12 | 24 |
| LSCO-850 | 285 | 317 (+32) | 335 (+50) | 346 (+61) | 362 (+77) |
| LSCO-875 | 275 | 307 (+32) | 320 (+45) | 328 (+53) | 334 (+64) |
| LSCO-900 | 265 | 288 (+23) | 300 (+35) | 306 (+41) | 312 (+47) |

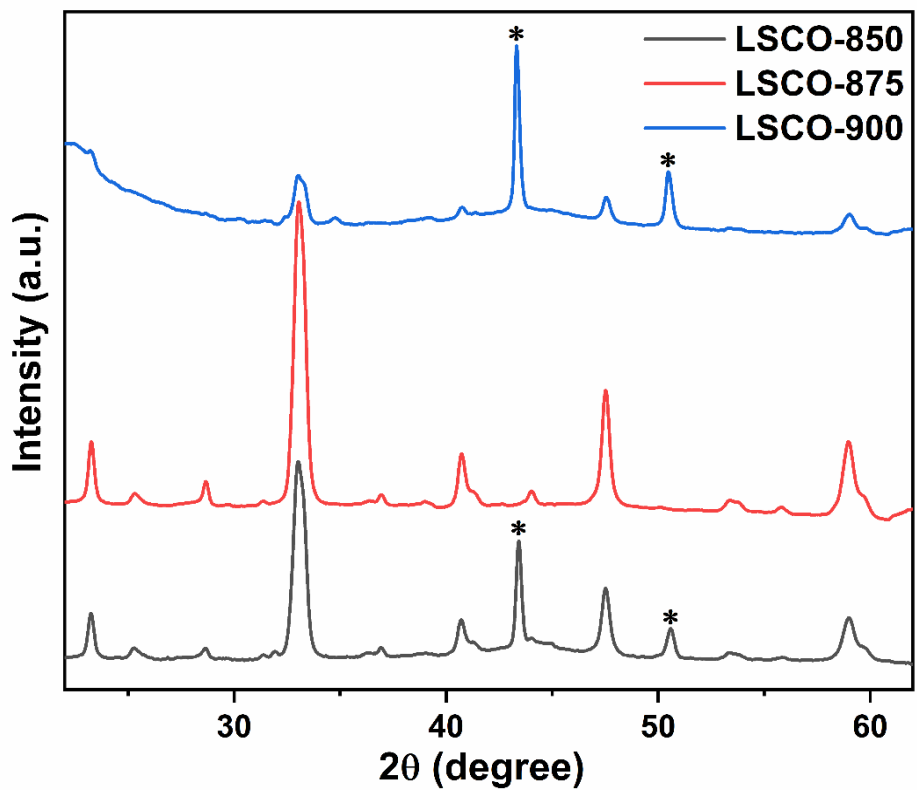

**Fig. S5**. The XRD results for the LSCO electrode after the CP test in 1 M KOH electrolyte (* diffraction reflexes of metallic copper caused by the copper holder of the electrode material).

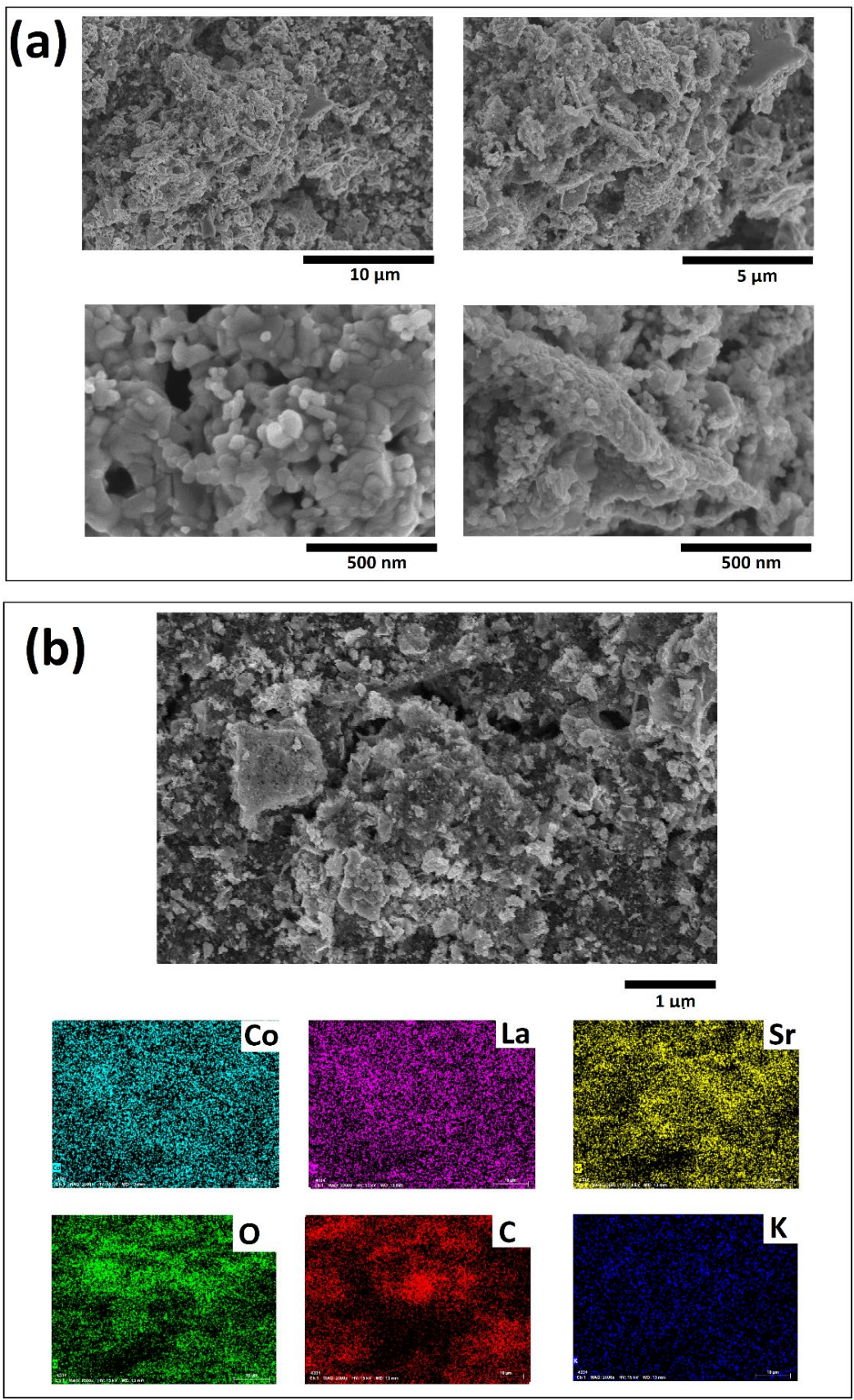

**Fig. S6**. SEM images and mapping of surface elements of the LSCO-850 electrode after long-term electrolysis.

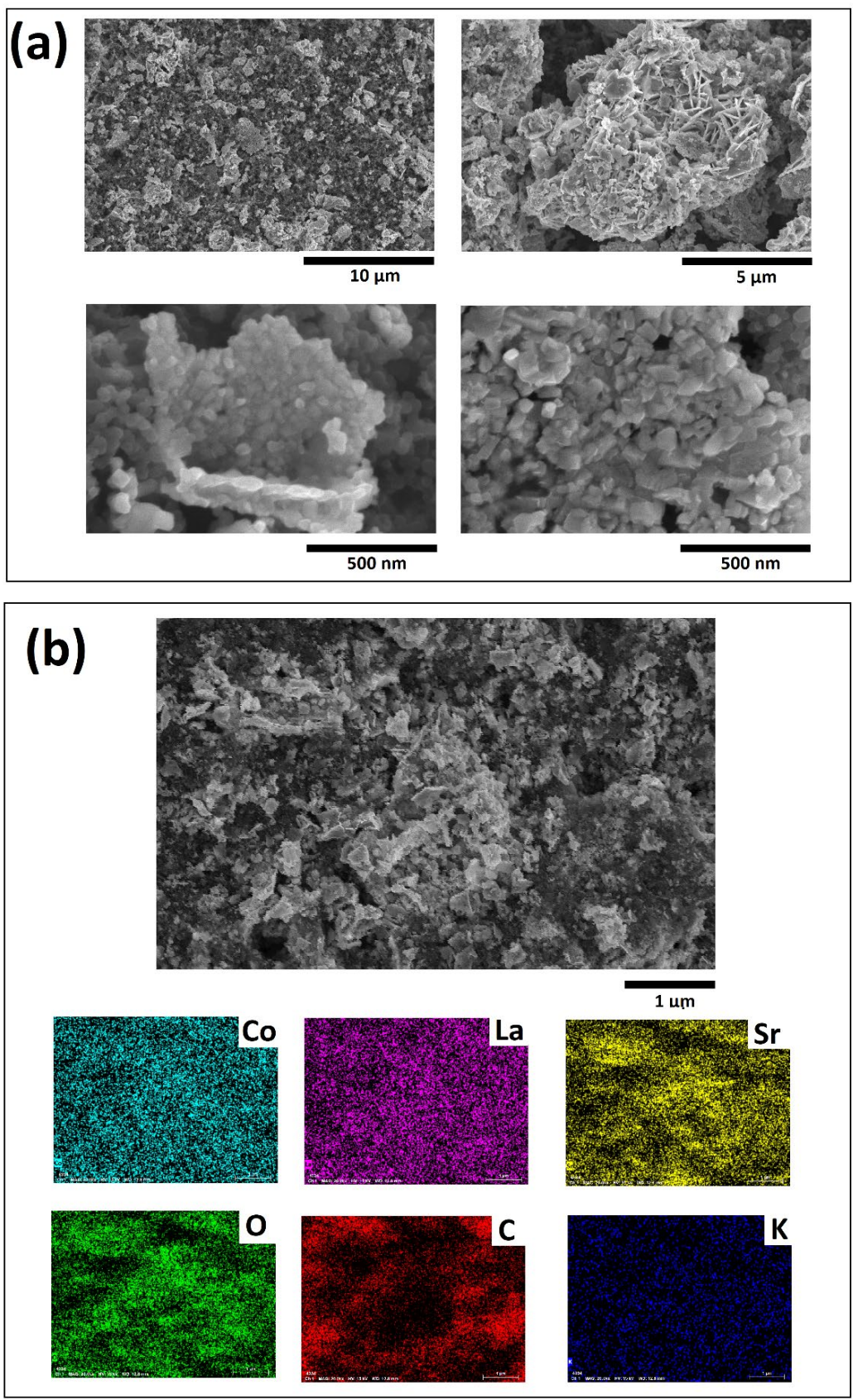

**Fig. S7**. SEM images and mapping of surface elements of the LSCO-875 electrode after long-term electrolysis.

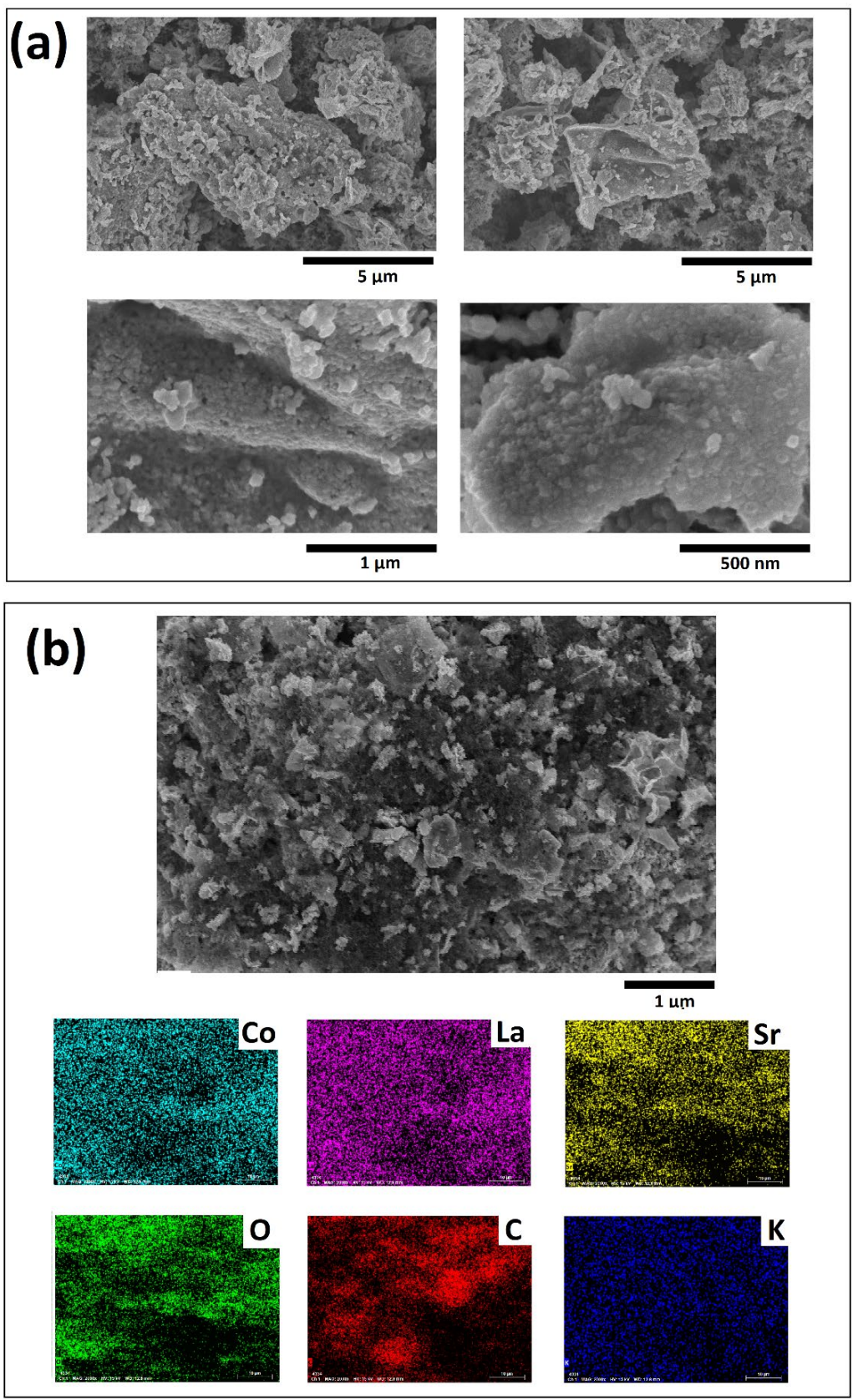

**Fig. S8.** SEM images and mapping of surface elements of the LSCO-900 electrode after long-term electrolysis.

## XPS of the LSCO-900 nanopowders

The survey spectra of the LSCO-900 samples before and after electrocatalysis show the existence of La, Sr, Co, and O peaks (Fig. S9(a)). At first glance, they seem very similar, but as shown below, an amorphous state of the LSCO-900 after electrocatalysis has been detected. In the high-resolution XPS spectra of the La3d, two major peaks, corresponding spin doublet La$3d_{5/2}$ and La$3d_{3/2}$, are observed (Fig. S9(b)). However, the spectra have a more complex profile due to charge transfer satellites, plasmon lines, and MNN Auger lines (marked as A and B in Fig. S9(b)). Similar results are also observed for the XPS spectrum of $La_2O_3$ [3] and La-containing manganite compounds [4]. The binding energy positions for the LSCO-900 before electrocatalysis are 832.9 eV (La$3d_{5/2}$) and 849.8 eV (La$3d_{3/2}$), with a difference of 16.8 eV which agrees with the standard deviation [5]. All La ions are in a trivalent state. Noteworthy, the XPS spectrum for the LSCO-900 after electrocatalysis is transformed to single-peaks with the binding energies of 833.3 (La$3d_{5/2}$) and 850.9 eV (La$3d_{3/2}$). It is associated with the amorphization of the LSCO-900 surface and the appearance of chemical defects due to a positive core-level shift of 0.4–0.9 eV [6]. All peak positions for the La, Sr, and O ions are listed in Table S4.

Fig. S9(c) shows Sr3d spectra decomposed to two main components from the perovskite lattice-bound Sr$_{lat}$ at lower binding energies (131.6 for $3d_{5/2}$ and 133.4 eV for $3d_{3/2}$) and from the surface-bound Sr$_{surf}$ at higher binding energies (133.5 for $3d_{5/2}$ and 135.3 eV for $3d_{3/2}$) for the LSCO-900 before electrocatalysis (see Table S4). The peak positions agree with other literature data [7, 8]. The surfaced Sr$_{surf}$ may be attributed to the formation of $SrCO_3$, SrO, and $Sr(OH)_2$ [9-11]. As it turned out, the Sr$_{surf}$ / Sr$_{lat}$ before and after electrochemical tests are 1.4 and 3, respectively. Moreover, a positive core-level shift of 1.9 eV is observed, showing the presence of oxygen vacancies. Additionally, after electrocatalysis, the three main peaks of Sr are transformed to somehow one single-peak. All this indicates an amorphization of the LSCO-900 surface as well.

As for O1s XPS before and after electrochemical tests (Fig. S9(d)), three components with energy positions of about 528.1–528.4 (O1), 530.0–531.2 (O2), and 531.2–533.8 (O3) eV can be distinguished in the spectra (see Table S4). Peak O1 should be attributed to the lattice oxygen, while components O2 and O3 are due to adsorbed oxygen and hydroxyl groups or water on the surface, respectively [6]. It should be noted that the ratio of $O_{ads}/O_{lat}$ before and after electrocatalysis increases by 2.3 and 2.5, respectively. Moreover, O3 contribution after electrocatalysis increases significantly, indicating growth in $H_2O$ molecules.

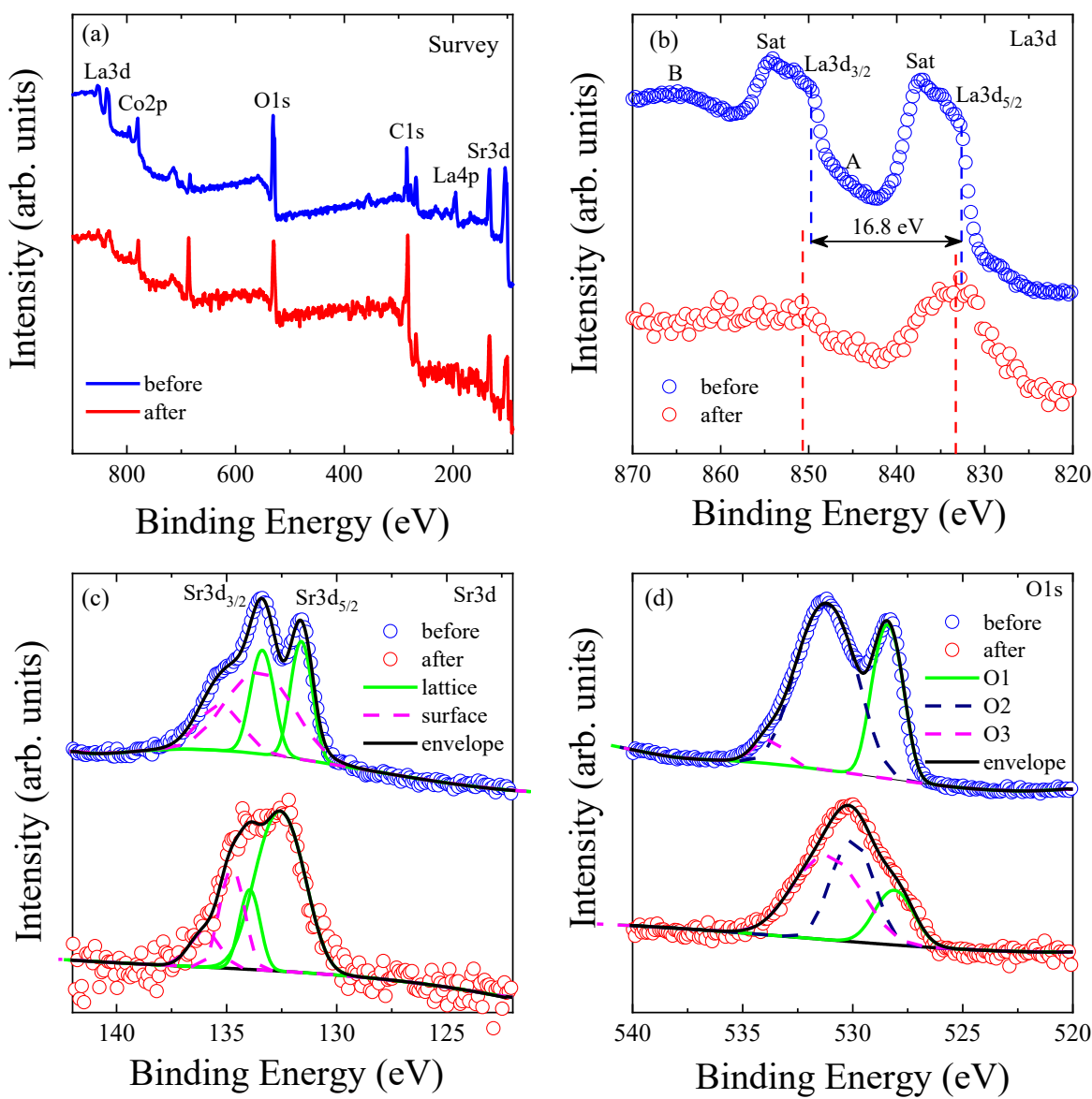

**Fig. S9.** XPS curves for the LSCO-900 samples before and after electrocatalysis: (a) survey spectra; (b) La3d spectra; (c) Sr3d spectra; and (d) O1s spectra.

**Table S4**

Energy positions of the La3d, Sr3d, and O1s X-ray photoelectron lines in the LSCO-900 samples before and after electrocatalysis.

| Electro-catalysis | Binding energy (eV) | | | | | | | | | | |
|---|---|---|---|---|---|---|---|---|---|---|---|
| | La3d | | | | Sr3d | | | | O1s | | |
| | La3d$_{5/2}$ | Sat | La3d$_{3/2}$ | Sat | Sr$_{lat}$5/2 | Sr$_{surf}$5/2 | Sr$_{lat}$3/2 | Sr$_{surf}$3/2 | O1 | O2 | O3 |
| Before | 832.9 | 837.4 | 849.8 | 854.3 | 131.6 | 133.4 | 133.5 | 135.3 | 528.4 | 531.2 | 533.8 |
| After | 833.3 | | 850.9 | | 132.6 | 133.9 | 134.8 | 136.0 | 528.1 | 530.0 | 531.2 |

**Table S5**

Energy positions of the Co2p, Co$^{2+}$, and Co$_3$O$_4$ X-ray photoelectron lines in the LSCO-900 samples before and after electrocatalysis.

| Electrocatalysis | Binding energy (eV) | | | | | |
|---|---|---|---|---|---|---|
| | Co2p$_{3/2}$ | | Co2p$_{1/2}$ | | Co$^{2+}$ | Co$_3$O$_4$ |
| | Co$^{3+}$ | Co$^{4+}$ | Co$^{3+}$ | Co$^{4+}$ | Sat | Sat |
| Before | 780.0 | 781.7 | 795.1 | 796.9 | 786 | 789.9 |
| After | 778.9 | 780.8 | 793.9 | 795.5 | – | 789.2 |